\documentclass[11pt]{article}
\usepackage{epsfig}
\usepackage{amsfonts}
\usepackage{latexsym}
\usepackage{epic}
\usepackage[colorlinks]{hyperref}
\newcommand{\ba}{\begin{array}}
\newcommand{\ea}{\end{array}}
\newcommand{\be}{\begin{equation}}
\newcommand{\ee}{\end{equation}}
\newcommand{\nn}{\nonumber}
\newcommand{\bea}{\begin{eqnarray}}
\newcommand{\ena}{\end{eqnarray}}
\newcommand{\beas}{\begin{eqnarray*}}
\newcommand{\enas}{\end{eqnarray*}}

\addtocounter{section}{0}
\renewcommand{\theequation}{\thesection.\arabic{equation}}
\def\u{{\bf{u}}}
\def\v{{\bf{v}}}
\def\w{{\bf{w}}}
\begin{document}

%\sloppy

\begin{center}
 {\LARGE On the solutions to the multi-parametric Yang-Baxter equations %:
}
 %{\large classification and alternative new free-fermionic solutions for $4\times 4$
% and $9\times 9$ $R$-matrices}
\end{center}

\begin{center}
{\bf
 Sh. Khachatryan\footnote{e-mail:{\sl shah@mail.yerphi.am}%; address:{ Alikhanian Br. str. 2, Yerevan 36, Armenia}
 }
 }
 \\ Yerevan
Physics Institute after A.~I.~Alikhanian, Alikhanian Br. str. 2, Yerevan 36, Armenia %\footnote{"National
%Scientific Laboratory after A.~I.~Alikhanian" Foundation},\\
%Alikhanian Br. str. 2, Yerevan 36, Armenia
\end{center}

\vspace{1.5 cm}
\begin{center}
{\LARGE
 Abstract}
\end{center}

 A unified approach is applied in the consideration of the
multi-parametric (colored) Yang-Baxter equations (YBE) and the
usual YBE with two-parametric R-matrices, relying on the existence
of the arbitrary functions in the general solutions. The
 colored YBE are
considered with the R-matrices defined on  two and three
dimensional states. We present an exhaustive study and the overall
solutions for the YBE with  $4 \times 4$ colored R-matrices. The
established classification includes new multi-parametric free
fermionic solutions. In the context of the given approach there
are obtained the colored solutions to the YBE with $9 \times 9$
R-matrices having 15 non-zero elements.

{\small\tableofcontents}

%\newpage

%\begin{verse}
  %text
 %\end{verse}

\section{Introduction}

 The Yang-Baxter equations (YBE), being formulated in the early works \cite{Y,Baxt78}
 (originated in \cite{Ons}),
 have crucial role in the theory of the integrable models
 in low dimensional statistical physics, quantum field theory and
 are the non-dividable parts of the theory of the quantum groups
 \cite{FT,Baxt,FTS,Z,Z1,grs}.
   Investigations of the
 solutions to YBE are still actual and are involved in newer fields
of the theoretical and mathematical physics.% \cite{LK}. % also in that context.
%Recently there is arisen an interesting application of the
%universal solutions
% to YBE to
% the theory of quantum computation \cite{LK}.

The usual quantum YBE are formulated as the system of the
equations
\bea
R_{ij}(u,v)R_{ik}(u,w)R_{jk}(v,w)=R_{jk}(v,w)R_{ik}(u,w)R_{ij}(u,v).\label{ybeuvw}
\ena
Here $R_{ij}(u,v)$ is $n^2\times n^2$ matrix, %, where n is the number of
%the values which the indexes $i,\; j$ acquire.
$u$ and $v$ are called spectral parameters. The matrix
$R_{ij}(u,v)$ is considered as an operator acting on the tensor
product of two $n$-dimensional states $V_i\otimes V_j$.
 The simplest and
well studied
 solutions to
 YBE are the $4\times
4$-matrices with eight non-zero entries. The symmetric solution is
presented in the paper \cite{Baxt}, which is the solution
corresponding to the 2d classical statistical eight-vertex model
(or the $XYZ$ Heisenberg model). This solution has difference
property - $R_{ij}(u,v)=R_{ij}(u-v)$, so the $R$-matrix is actually
one-parametric, and the corresponding YBE can be presented as
\bea
R_{ij}(u)R_{ik}(u+w)R_{jk}(w)=R_{jk}(w)R_{ik}(u+w)R_{ij}(u).\label{ybeu-w}
\ena
 In
the context of the $1+1$ quantum field theory \cite{Z,Z1}, where
the $R$-matrix plays the role of the scattering matrix of two
particles, the YBE just ensures the factorization of the
many-particle scattering matrices into the products of
two-particle scattering matrices. Here the spectral parameters
are simply the rapidities of
the relativistic particles, and the difference property reflects %the isotropy
%and homogeneity of the space.
the relativistic invariance of the system. Besides of the rapidity
the scattering matrix can be depended also from the extra %inner
characteristics of the particles - "colors". In the papers
\cite{Felder,BVS,RKS} the multi-parametric  YBE are studied with
colored $R$-matrices $R_{ij}(u;p,q)$ (the color parameters $p$ and
$q$ are attached to the the $i$-th and $j$-th spaces respectively)
\bea
R_{ij}(u;p,q)R_{ik}(u+w;p,r)R_{jk}(w;q,r)=R_{jk}(w;q,r)R_{ik}(u+w;p,r)R_{ij}(u;p,q).\label{ybeupq}
\ena
The corresponding solutions have the so called free-fermionic
property \cite{WuChan},
\bea
R_{00}^{00}R_{11}^{11}+R_{01}^{01}R_{10}^{10}-R_{01}^{10}R_{10}^{01}-R_{00}^{11}R_{11}^{00}=0,
\ena
which means that the corresponding 1d quantum theories can be
presented by means of the scalar free fermionic chains.

In the  paper \cite{SSh} there is discussed a two-parametric
solution, which is just the solution to the YBE with $R$-matrices
having two pairs of rapidities, so that for each pair the
$R$-matrix has difference property - $R(u,v;u',v')\equiv
R(u-v,u'-v')$. Then YBE has the following form
\bea R_{ij}(u,u')R_{ik}(u+w,u'+w')R_{jk}(w,w')=
R_{jk}(w,w')R_{ik}(u+w,u'+w')R_{ij}(u,u'). \label{ybeuu'}\ena

In the recent paper \cite{shkh} there are presented
multi-parametric solutions, which being free-fermionic, do not
correspond to the mentioned cases. The new $4\times 4$ solutions
\cite{shkh} (see the subsection 4.4 therein) are formulated as two-parametric
 solutions, which in general case have no difference
property and contain  arbitrary functions.

The one of the purposes of this paper is to fill the gap, which
exists in the study of the multi-parametric systems of YBE  and in the
hierarchy of the free-fermionic solutions for the case $n=2$. Then
we show that the YBE put the restriction on the number of the
possible "colors", which can be "visible" in the solutions for the
given $n$. For analyzing the set of YBE we  propose and apply
 a straightforward way,
formulated precisely in section 3. We classify here the all types
of the multi-parametric solutions for the matrices of  the
general eight-vertex
kind, analyzing YBE with the $4\times 4$-matrices.
Meanwhile the whole approach could be valid also for the cases
with higher dimensional matrices, and  particularly, we consider
 the YBE with $n=3$ too and derive a family of multi-parametric
$R$-matrices. This set of the solutions entirely describes the
situation when the $R$-matrices have $15$ non-zero elements. The
extensions of the solutions to the cases with the matrices having
more non-zero elements will be presented further. In particular
cases the obtained solutions
 correspond to the known three-parametric colored solutions,
the eight-vertex model's solution and the other already observed
YBE solutions both for  $n=2,3$ \cite{Baxt,
Felder,BVS,RKS,shkh,GU,JKL,BS,PerkSch,PerkAu-Yang,33,33osp}.

The paper is organized in the following way. In the Section 2 the
YBE is presented in general multi-parametric formulation. In the
subsections 2.1 and 2.2 the particular cases are discussed,
corresponding to the situations $b_i=0$ and $d_i=0$. In the
subsection 2.3 the general case is discussed. There the
generalized multi-parametric version of the eight-vertex model's
matrix is obtained, and we see that beside of these solutions  the all
remaining solutions  have free-fermionic property. We are
presenting the general free-fermionic solutions as matrices with
two arbitrary functions (it corresponds to the
four-parametric solutions) and two arbitrary constants. When one
of the constants vanishes, the solution coincides with the known
colored solutions \cite{Felder,BVS}. The $sl_q(2)$-invariant
solution in \cite{shkh} at $q=i$ for the case $c_{1,2}=0$ (the
eigenvalues of the quadratic Casimir operator of the
representations spaces on which $R$-matrix acts), belongs to the
family of the four-parametric solutions  when there are two
arbitrary constants and one arbitrary function, while the other
function is parameterized by exponential function in a specific
way
 imposed by the condition of the algebra invariance.
 In the Section 3 we present all the
results and conclusions followed from the performed analysis.
Therein a table is presented with the classification of the all
principal types of the $4\times 4$ solutions. In the context of
the given approach in the next section we have solved
multi-parametric YBE for  %definite
general colored $9\times 9$ $R$-matrices with $15$ non-zero
elements. %New solutions are obtained.
%The more complete analysis for this case will
% be presented further.
The first part of the Appendix  is
devoted to the detailed discussion of the analysis of  YBE %solutions
for a particular case typical for the general solutions with
arbitrary functions,
 and in the second part  the main different types of the $4\times 4$-solutions are
presented in apparent matrix formulations. %In the subsequent part of the
%appendix a table is presented with the classification of the all principal
%types of the $4\times 4$ $R$-matrices.

\section{
%system and general conditions
General multiparametric solutions to YBE with $4\times 4$
$R$-matrices} \addtocounter{section}{0}\setcounter{equation}{0}
%{\color{red}{$\ast\ast\ast\ast\ast\ast\ast\ast\ast\ast\ast\ast\ast\ast\ast\ast\ast$}
%\color{blue}{$\ast\ast\ast\ast\ast\ast\ast\ast\ast\ast\ast\ast\ast\ast\ast\ast\ast$}
%\textcolor[rgb]{1.00,0.44,0.18}{$\ast\ast\ast\ast\ast\ast\ast\ast\ast\ast\ast\ast\ast\ast\ast\ast\ast$}}

The general form of the quantum Yang-Baxter equations with
multi-parameters %let us write
can be presented as
\be
R_{ij}(\mathbf{u},\mathbf{v})R_{ik}(\mathbf{u},\mathbf{w})R_{jk}(\mathbf{v},\mathbf{w})=
R_{jk}(\mathbf{v},\mathbf{w})R_{ik}(\mathbf{u},\mathbf{w})R_{ij}(\mathbf{u},\mathbf{v}).
\label{ybe}\ee
Here under the spectral parameters $\mathbf{u},\; \mathbf{v},\;
\mathbf{w}$, written in the "bold" shrift, we mean the all
possible set of the parameters, which the $R$ matrix  can acquire
- $\mathbf{u}=\{u_1,u_2,...\},\; \mathbf{v}=\{v_1,v_2,...\}\;
\mathbf{w}=\{w_1,w_2,...\}$. As the matrix acts on the tensor
product of two vector spaces, it
%conditions
means that the sets of the spectral parameters are attached
to the corresponding vector spaces. %In the particle scattering's
%theory in $1+1$ space-time, where the YBE just ensures the
%factorization of the many-particle scattering matrices, this
%property arises naturally, as the spectral parameters are simply
%the rapidities of the scattering particles. The homogeneity and
%isotropy of the space imply that the matrix depends on the
%difference of the rapidities $R_{ij}(u_i,u_j)=R_{ij}(u_i-u_j)$.

Let us explore the matrices having the property of "particle
number" conservation by mod(2) ($\mathcal{Z}_2$ grading symmetry).
In matrix representation it means $ R_{ij}^{kr}\neq 0$ if
$i+j+k+r=0 \; (mod\; 2)$.
If the indexes $i,j...$ take only two values $0,\; 1$, then $R$ has the
following $4\times 4$ matrix form
 \bea \label{rg}&R(\u,\w)=\left(\ba{cccc}
R_{00}^{00}&0&0&R_{00}^{11}\\
0&R_{01}^{01}&R_{01}^{10}&0\\
0&R_{10}^{01}&R_{10}^{10}&0\\
R_{11}^{00}&0&0&R_{11}^{11}\\
\ea\right)%(\u,\w)
\equiv %&\\  &
\left(\ba{cccc}
a_1(\u,\w)&0&0&d_1(\u,\w)\\
0&b_1(\u,\w)&c_1(\u,\w)&0\\
0&c_2(\u,\w)&b_2(\u,\w)&0\\
d_2(\u,\w)&0&0&a_2(\u,\w)\\
\ea\right)&\label{rx}
%(u-v),
\ena
which is similar to the $R$-matrix of the eight-vertex model. We %shall
use the commonly %following %usually
adopted notations (\ref{rx}) for the matrix elements.
%
% \bea \nn
%&R^{00}_{00}=a_1,\; R^{11}_{11}=a_2,\;
%R^{01}_{01}=b_1,\;R^{10}_{10}=b_2,\;&\\
% &R^{10}_{01}=c_1,\; R^{01}_{10}=c_2,\;
% R^{11}_{00}=d_1,\; R^{00}_{11}=d_2.&
%\label{rx}\ena
%

%At first
We are considering the complete and  motivated cases,
with $a_i\neq 0,\;c_i\neq 0$ and %(below $\check{R}=P R$, where $P$
%is a permutation operator changing the positions of the states,
%and $\mathbf{I}$ is unit operator)
%
\bea \check{R}(\u,\u) =\mathbf{I}, \label{unite}\ena
where $\check{R}=P R$, $P$ is a permutation operator changing the
positions of the states, and $\mathbf{I}$ is the unit operator.

 The simplest equations followed from YBE (\ref{ybe}) are
\bea &c_1({\u}_1,\u_2)c_1(\u_2,\u_3) c_2(\u_1,\u_3) - c_1(\u_1,\u_3) c_2(\u_1,\u_2) c_2(\u_2,\u_3)=0,&\label{eq1}\\
&c_1(\u_1,\u_3) d_1(\u_2,\u_3) d_2(\u_1,\u_2) - c_2(\u_1,\u_3) d_1(\u_1,\u_2) d_2(\u_2,\u_3)=0,&\label{eq2}\\
& c_1(\u_1,\u_2) d_1(\u_2,\u_3) d_2(\u_1,\u_3) - c_2(\u)
d_1(\u_1,\u_3) d_2(\u_2,\u_3)=0.&\label{eq3}\ena
The general solutions to them we can parameterize by means of an
arbitrary function $g(\u)$ and an arbitrary constant $d_0$
\bea \frac{c_2(\u,\w)}{c_1(\u,\w)}=\frac{g(\u)}{g(\w)},\quad
\frac{d_2(\u,\w)}{d_1(\u,\w)}=d_0 g(\u)g(\w). \ena
%
%Let us recall once again, that the parameters $u,\;w$ are
%synthesized, joint parameters with meaning of the sets of the
%parameters $\{u\},\;\{w\}$, connected with given states. % and not
%just single ones.

As it is known there is a possibility to redefine the matrix
elements of the matrix $R_{ij}$ by the following transformations
\cite{grs}
\bea R_{n_i n_j}^{p_i p_j}(\u_i,\u_j)\Rightarrow R_{n_i n_j}^{p_i
p_j}(\u_i,\u_j)\frac{\mathrm{f}_{n_i}(\u_i)\mathrm{f}_{n_j}(\u_j)}
{\mathrm{f}_{p_i}(\u_i)\mathrm{f}_{p_j}(\u_j)}, \label{auf} \ena
induced from the following %transformations
change of the vector basis $\mathrm{e}_{n_{i,j}}$ ($n_{i,j}=0,1$,
$p_{i,j}=0,1$) of the space $V_{i,j}$ - $\mathrm{e}_{n_{i,j}}\to
\mathrm{f}_{n_{i,j}}\mathrm{e}_{n_{i,j}}$ . The transformations
affect only the elements $R_{01}^{10}$, $R_{10}^{01}$,
$R^{00}_{11}$, $R_{00}^{11}$. Taking the functions in this way
$\left(\frac{\mathrm{f}_{0}(\u)}{\mathrm{f}_{1}(\u)}\right)^2=
\sqrt{d_0}g(\u)$, we can make the elements $c(d)_i$, $i=1,2$
equal. So, we can consider % take
  $c_1=c_2$ and
$d_1=d_2$ afterwards.
%This is the case taken in the papers \cite{BVS,Felder}.

We shall explore all the possible cases in detail.

We can set $c_1=c_2=1$ taking into account the normalization
freedom of the $R$-matrix.
 The set of the independent equations of YBE is brought
in the Appendix, (\ref{ybe1c}-\ref{ybe1e}) (the conditions
(\ref{condit}) ensue from the YBE for each discussed case, as it will be
shown).

\subsection{$a_1=a_2$, $b_i=0$, $i=1,2$}

 %The relations $b_i=0$ satisfy to (\ref{re}). Then
 In this subsection we are demonstrating a plain case with the conditions
  $a_1=a_2$, $b_i=0$, $i=1,2$, for which the
extra colored parameters are actually absent,
 and can be introduced only by taking into account the transformation
 freedom (\ref{auf}).
  From
the analysis of the whole set of the equations  the following
general solution follows

\bea \label{yr1}&a_1(\u,\w)=a_2(\u,\w)=\frac{a(\u)a(\w)}{1+(d(\u)-d_0)d(\w)},&\\
\label{yr2}&d_1(\u,\w)=d_2(\u,\w)=\frac{d(\u)-d(\w)}{1+(d(\u)-d_0)d(\w)},&\\
\label{yr3}&a(\u)^2-d(\u)^2-1+d(\u)d_0=0.& \ena
The first two equations are just the relations expressing the full
functions $f_i(\u,\w)$, $f=a,d$ via the elementary functions
$f_i(\u)$.  We see, that the functions $a_i(\u)$ and $d(\u)$ are
defined so, that $a_i(\u)=a_i(\u,\u_0)$ and $d(u)=d(\u,\u_0)$, if
at the point $\u_0$ we have $a(\u_0)=1$, $d(\u_0)=0$. Note, that
$\u=\w$ corresponds to the normalization point
$a(\u,\u)=1,\;d(\u,\u)=0$. Further we can fix
$\u_0=\mathbf{0}=\{0,0,...\}$ symbolically. The last equation
(\ref{yr3}) is the consistency condition of the YBE. This
relation is true also for the functions $a_i(\u,\w)$ and
$d_i(\u,\w)$. Any $R$-matrix having the elements (\ref{yr1},
\ref{yr2}) with arbitrary function $d(\u)$ and with $a(\u)=\pm
\sqrt{d(\u)^2+1-d(\u)d_0}$ is YBE solution.  As here
 we have only one arbitrary
function, then the dependence from the multi-spectral parameters
$\u$ one can encode in the argument of that function. We see, that
the consistency condition implies trigonometric
parameterization. %, if the functions have only one arguments.
% A simple parameterization could be
 %
% \bea \label{ras}
%a(\varepsilon)=\frac{\cosh{[\varepsilon+u_0]}}{\cosh{[u_0]}},\quad
%d(\varepsilon)=\frac{\sinh{[\varepsilon+u_0]}}{\cosh{[u_0]}}-\tanh{[u_0]},\quad
%d_0=-2 \tanh{[u_0]}.
 %\ena
%
%With this parameterization the functions
%$a_i(\varepsilon,\varepsilon'),\; d(\varepsilon,\varepsilon')$,
%defined from the relations (\ref{yr1}, \ref{yr2}), look like
%two-parametric functions.
 If to impose the constraints
$a_i(u,w)=a_i(u-w)$ and $d_i(u,w)=d_i(u-w)$, then the relations
(\ref{yr1}, \ref{yr2}) become functional equations on the
corresponding functions and the solution is the following
 \bea \label{rtar}
a(u)=\frac{\sinh{[u_0]}}{\sinh{[u+u_0]}},\quad
d(u)=\frac{\sinh{[u]}}{\sinh{[u+u_0]}},\quad d_0=2 \cosh{[u_0]},
 \ena
which is the solution \cite{SSh} describing the XYZ-model with the
coupling parameters $J_x=-J_y$, $J_z=\cosh{[u_0]}$.

%One can combine these two parameterizations in order to obtain
%more general solution
%$R_{ij}(\varepsilon_i,\varepsilon_j;u_i,u_j)$.
 Note, that the
restriction that the dependence of the matrix elements from the
spectral parameters must be in difference form $R_{ij}(u-w)$ fixes
the functions $a(\u),\; d(\u)$,  meanwhile in general these two
functions are connected by one relation (\ref{yr3}). Hence one of
the functions, say  $a(\u)$,
is arbitrary and we can choose any parameterization for it. %As
%example,  a plain parameterization could be
 %
% \bea \label{ras}
%$a(\varepsilon)=\frac{\cosh{[\varepsilon+\varepsilon_0]}}{\cosh{[\varepsilon_0]}},\quad
%d(\varepsilon)=\frac{\sinh{[\varepsilon+\varepsilon_0]}}{\cosh{[\varepsilon_0]}}
%-\tanh{[\varepsilon_0]},\quad d_0=-2 \tanh{[\varepsilon_0]}$.
 %\ena
%With this parameterization the functions
%$a_i(\varepsilon,\varepsilon'),\; d(\varepsilon,\varepsilon')$,
%defined from the relations (\ref{yr1}, \ref{yr2}), look like
%two-parametric functions. bring the dependence from $\varepsilon, \;\varepsilon'$ to the
%form similar to (\ref{rtar}).
 However by means of definite
%reparameterizations $\varepsilon\to u$, $\varepsilon'\to w$ we can  bring the dependence from $\varepsilon,
%\;\varepsilon'$ to the form similar to (\ref{rtar}).
 parameterization (\ref{rtar}) %. %, so the above functions will have the
%simple form $R_{ij}(u-w)$.
%In this way any two-parametric solution with the given
any two-parametric solution $R_{ij}(a(\u),a(\w))$ with the
conditions (\ref{yr1}-\ref{yr3}) can be brought to the actually
one-parametric form $R_{ij}(u-w)$.

%So, the
%appearance of the matrix elements of
%$R_{ij}(\varepsilon_i,\varepsilon_j;u_i-u_j)$ can be different due
%to the dependence from the variables $\varepsilon$. If to choose
%$R_{ij}(\varepsilon,\varepsilon;0)=I$ then the matrix
%$R_{ij}(\varepsilon,\varepsilon;u_i-u_j)$ will coincide with the
%solution (\ref{rtar}). Then we can claim that
%$R_{ij}(\varepsilon_i,\varepsilon_j;0)$ do coincide with the
%solution, say brought in (\ref{ras}). The resulting function s
%will be
%
 %\bea \label{rasfk}
%a(\varepsilon)=\frac{\cosh{[\varepsilon+u_0]}}{\cosh{[u_0]}},\quad
%d(\varepsilon)=\frac{\sinh{[\varepsilon+u_0]}}{\cosh{[u_0]}}-\tanh{[u_0]},\quad
%d_0=2 \cosh{[u_0]}.
% \ena
%

\subsection{$d_i=0$, $i=1,2$}

%The more complete solution is derived without any restrictions on
%the functions $a_1,\; a_2$ and $b_1,\;b_2$. Then,
We have $a_i(\u,\u)=1$, $b_i(\u,\u)=0$ from the relation
(\ref{unite}). Let us again take a symbolic fixed point
$\u=\mathbf{0}$, and set $a_i(\u)=a_i(\u,\mathbf{0})$,
$b_i(\u)=b_i(\u,\mathbf{0})$. The YB equations give the following
expressions for the complete functions $f_i(\u,\w)$ by means of
the elementary functions $f_i(\u)=f_i(\u, \mathbf{0})$
\bea &a_1(\u,\w)=b_1(\w)b_2(\u)+\frac{a_1(\u)}{a_1(\w)}(1-b_1(\w)b_2(\w)),&\nn\\
&a_2(\u,\w)=a_1(\w)a_2(\u)+\frac{b_2(\w)}{b_2(\u)}(1-a_1(\u)a_2(\u)),&\label{ruw}\\
&b_1(\u,\w)=a_2(\w)b_1(\u)-a_2(\u)b_1(\w),\quad
b_2(\u,\w)=a_1(\w)b_2(\u)-a_1(\u)b_2(\w).&\nn
 \ena
The all relations between the functions $a_i(u),\; b_i(u)$, coming
from the YBE, can be expressed by the following constraints of
familiar type
\bea &\frac{a_1(\u)a_2(\u)+b_1(\u)b_2(\u)-1}{b_1(\u) a_1(\u)}=\Delta&\label{abc}\\
&\frac{a_2(\u)b_2(\u)}{a_1(\u)b_1(\u)}=k.& \label{abx}\ena
The relations (\ref{abc}, \ref{abx}) in general case are not valid
for the full matrix elements $f_i(\u,\w)$. Here the parameters
$\Delta,\; k$ are constants, and the second equation (\ref{abx})
must take place when $\Delta\neq 0$ (otherwise it is not necessary
condition). Indeed, one could also immediately  this
analyzing the YBE in the common way \cite{Baxt}, considering the
equations (\ref{ybe}) as homogeneous linear equations in respect
of the functions $f_i(\u,\v)$ (or either $f_i(\u,w)$ or
$f_i(\v,\w)$). Then the relations on the matrix elements are
arisen from the consistency conditions of the homogeneous
equations (formulated as vanishing of the determinants of the
matrices composed by the corresponding coefficient functions).
 Taking now the YBE as homogeneous linear equations in respect
of the functions depending, say on the parameters $(\v,\w)$, and
writing down the consistency conditions of the equations as
\bea \nn & \Big(a_1(\u,\v) a_2(\u,\w) b_1(\u,\v) b_2(\u,\w) -
a_1(\u,\w) a_2(\u,\v) b_1(\u,\w) b_2(\u,\v)\Big)&\\
&\times\Big(a_1(\u,\v) a_2(\u,\v) + b_1(\u,\v) b_2(\u,\v) -
1\Big)=0,&\label{re}\ena
we see that when the free fermionic condition takes place (the
second row in (\ref{re}) vanishes), then it is not necessity for
the  vanishing of the first row.

\paragraph{$\bullet$ $\Delta\neq 0$. }

In general the solutions contain two arbitrary functions, say
$a_2(\u)$ and $b_1(\u)$ (the other two functions, $a_1(\u)$ and
$b_2(\u)$, can be obtained from the constraints (\ref{abc},
\ref{abx})) and two arbitrary parameters $\Delta$ and $k$
\bea& a_1(\u)=\frac{a_2(\u)}{a_2(\u)^2-\Delta a_2(\u)b_1 (\u)+k
b_1(\u)^2},\quad b_2(\u)=\frac{k b_1(\u)}{a_2(\u)^2-\Delta
a_2(\u)b_1 (\u)+k b_1(\u)^2}.&\label{ab_abc}\ena

As it was noted, we can encode dependence from two arbitrary
functions into the arguments - %the dependence of
two sets of independent spectral parameters ($\u=\{u,p\},\;
\w=\{w,q\}$), $f_i(\u,\w)=f_i(u,w;p,q)$. The choice of the
appropriate parameterization, in which the dependence from some
variables would have difference property, say
$R(u,w;p,q)=R(u-w;p,q)$, brings to differential equations. Taking
then  $\textbf{0}=\{0,0\}$, we shall have
$f_i(u,w;p,0)=f_i(u-w,p)$, where %as in the previous notations
$f_i(\u)=f_i(\u,0)\equiv f_i(u,0;p,0)=f_i(u;p)$. Combining the
relations (\ref{ruw}, \ref{abc}, \ref{abx}), and expanding the
functions $f_i(u,w;p,0)$ near the point $w=0$, we shall come to
$f_i''(u;p)\approx f_i(u,;p)$ (the second differential is taken
over the variable $u$). It means that the parameterization must be
taken by trigonometric functions over the variables $u,\; w$. So,
the functions will have the form $f_i(u;p)=g_i(p)\sinh{[u+u_i]}$,
with appropriate chosen functions $g_i(p)$ and constants $u_i$,
depending on $k,\;\Delta$. From another hand the desired
parameterization one could obtain in this way. We can see that
the functions $g(\u)$
and $g(\w)$, with $g(\u)^2=a_2(\u)^2-\Delta a_2(\u)b_1 (\u)+k
b_1(\u)^2$, are factorized in the expressions of the functions
$f_i(\u,\w)$. Taking
$a_2(\u)=g(\u)\sinh{[\phi(\u)+\phi_0]}/\sinh{[\phi_0]}$, $b_1(\u)=
\sinh{[\u]} g(\u)/(c_0 \sinh{\phi_0})$, with
$(\sinh{[\phi_0]})^2=(\Delta^2-4 k)/(4 k)$ and $c_0=\sqrt{k}$
(i.e. $\Delta=2\cosh[\phi_0] c_0$), the solution can be written as
\bea R(\u,\w)\!\!\!=\!\!\!\left(\!\!\!\ba{cccc}
 \frac{g(\w)\sinh{\![\phi(\u)-\phi(\w)+\phi_0]}}{g(\u) \sinh{\![\phi_0]}}&0&0&0\\
0& \frac{ g(\u)g(\w)\sinh{\![\phi(\u)-\phi(\w)]}}{c_0 \sinh{\![\phi_0]}}&1&0\\
0&1& \frac{c_0\sinh{\![\phi(\u)-\phi(\w)]}}{g(\u)g(\w)\sinh{\![\phi_0]}}&0\\
0&0&0&\frac{g(\u)\sinh{\![\phi(\u)-\phi(\w)+\phi_0]}}{g(\w)
\sinh{\![\phi_0]}}\ea\!\!\!\right).\nn\\
 \ena
%
%where the arguments of the functions are presented in such a way,
%which shows that this is a multiparametric solution, and we can
 The arguments in the previous discussion correspond to
   $\u=\{u,p\}$, $\w=\{w,q\}$, and $\phi(\u)=u$, $g(\u)=p$,
$\phi(\w)=w$, $g(\w)=q$. If to take into account
 also the parameters followed from the transformations (\ref{auf}),
 then after the following
 notations
\be\label{fst} \frac{\mathrm{f}_1(\u)}{\mathrm{f}_0(\u)}\equiv
t,\quad \frac{\mathrm{f}_1(\w)}{\mathrm{f}_0(\w)}\equiv s, \ee
we can write the $R(\u,\w)$-matrix as follows (now $\u=\{u,p,t\}$
and $\w=\{w,q,s\}$ and $\mathbf{0}=\{0,1,1\}$)
\bea R(u-w;p,q;s/t)=\left(\ba{cccc}
 \frac{q\sin{(u-w+u_0)}}{p \sin{u_0}}&0&0&0\\
0& \frac{p\; q \sin{(u-w)}}{c_0\sin{u_0}}&\frac{t}{s}&0\\
0&\frac{s}{t}& \frac{c_0\sin{(u-w)}}{p\; q \sin{u_0}}&0\\
0&0&0&\frac{p\sin{(u-w+u_0)}}{q \sin{u_0}}\ea\right)\label{rpqst}
\ena
Of course, also another choices are possible for parameterization,
say $\phi(\u)=u $, $g(\u)=e^{u} (p+u)$,
$\frac{\mathrm{f}_1(\u)}{\mathrm{f}_0(\u)}=e^{\alpha u} t$, and so
on. The dependence from the arbitrary constant $c_0$ can be
 eliminated by redefinition $g(\u)\to \sqrt{c_0} g(\u)$ ($p\to \sqrt{c_0} p,\;
 q\to \sqrt{c_0} q$). From the
 ordinary $XXZ$ the parameterization (\ref{rpqst}) differs by transformations
  of
 the basis states
 similar to the discussed one (\ref{auf}), if to redefine the matrix elements,
 as it was in the paper \cite{SSh} for the solution (5.25)
  $r_{X_{-}X Z}(u)$ -
${R}_{ij}^{i'j'}\to R_{i\;\bar{j}}^{i'\;\bar{j}'}$, where
$\bar{i}=mod[i+1]2$, (i.e. it corresponds to the transformations (\ref{auf})
after interchanging the indexes $0$ and
$1$).

The generalized relations similar to the eq.s (\ref{abc}) and
(\ref{abx}) for the complete functions $f_i(\u,\w)$ are the
following ones
\bea \frac{a_1(\u,w)a_2(\u,\w)+b_1(\u,\w)b_2(\u,\w)-1}{2\sqrt{
a_1(\u,\w)b_1(\u,\w)a_2(\u,\w)b_2(\u,\w)}
}=\cosh{[\phi_0]}\label{abce}\\
\frac{a_2(\u,\w)b_2(\u,\w)}{a_1(\u,\w)b_1(\u,\w)}=k/g(\w)^4.
\label{abxe} \ena

\paragraph{$\bullet$ $\Delta=0$.}

Here the first relation in (\ref{abc}) is enough for the functions
in (\ref{ruw}) with arbitrary  $a_i(\u),\; b_i(\u)$,
connected by one constraint, to be solutions to YBE.  The matrix
elements of $\check{R}(\u,\w)$, taken into account the consistency
conditions, can be written as
\bea &a_1(\u,\w)=a_1(\u)a_2(\w)+b_1(\w)b_2(\u),\quad
a_2(\u,\w)=a_1(\w,\u)&\\
& b_1(\u,\w)=b_1(\u)a_2(\w)-b_1(\w)a_2(\u), \quad
b_2(\u,\w)=b_2(\u)a_1(\w)-b_2(\w)a_1(\u),&\\
&1-a_1(\u)a_2(\u)-b_1(\u)b_2(\u)=0.&\label{a12} \ena
 As in the previous discussion, we can  interpret the solutions in two equivalent ways. We can
 say, that we have three independent and arbitrary functions (e.g. $a_1,\; a_2,\;b_1$)
 in the solution, or we can say, that we have a solution with
 six parameters $\{a_1(u),\; a_2(u),\;b_1(u);\; a_1(w),\;
 a_2(w),\;b_1(w)\}$ (three pairs of independent parameters).
 {\it In general the two-parametric solution with $n$ arbitrary functions   is equivalent to the
 $2n$-parametric solution.}
  Writing %hen we write
  the functions in terms of the
 composite parameters $\u,\;\w$ we unify and extend two
 interpretations.

 The demand
that the dependence from the spectral parameters to be in
difference form (with one spectral parameter) brings the solutions
to the known cases, which are the $XX$-model's matrix
($b_i(u)=\sin{u}$, $a_i(u)=\cos{u}$) or the matrix of the
$XX$-model in transverse magnetic field, with $b_2(u)=b_0
b_1(u)=\sin{u}/\sin{u_0}$ and
$(a_1(u)-a_2(u))/b_1(u)=\mathrm{constant}\equiv 2\cos{u_0}$,
$a_1=\sin{(u+u_0)}/\sin{u_0}$, $a_2(u)=\sin{(u_0-u)}/\sin{u_0}$.
When $u_0=\pi/2$ the second case corresponds to the first one. In
general  we can take as example
\bea
a_1(\u)=f(\u)\frac{\sin{[\phi(\u)+\phi_0]}}{\sin{[\phi_0]}},\quad
b_1(\u)=g(\u)\frac{\sin{[\phi(\u)]}}{\sin{[\phi_0]}},\\
a_2(\u)=\frac{1}{f(\u)}\frac{\sin{[\phi_0-\phi(\u)]}}{\sin{[\phi_0]}},\quad
b_2(\u)=\frac{1}{g(\u)}\frac{\sin{[\phi(\u)]}}{\sin{[\phi_0]}},\label{parfg}
\ena
with arbitrary multiparametric functions
$f(\u),\;g(\u),\;\phi(\u)$ and arbitrary constant $\phi_0$. Note,
that in the unique constraint (\ref{a12}) there is no any
constant, and hence $\phi_0$ is not a relevant constant and can be
eliminated by appropriate reparameterizations, and the same is
valid also for the constant $u_0$ introduced below in
(\ref{a_fr}).

 With the parameterization (\ref{parfg}) the independent functions
 are expressed by the arbitrary functions $\phi(\u),\; f(\u),\;g(\u)$.
 When
$f(\u)=1/g(\u)$ the $\check{R}$-matrix in terms of the
argument-function $\phi(\u)$ acquires difference property. If to
set the composite arguments consisting of three parameters,
$\u=\{u,p,\bar{p}\}$ and $\w=\{w,q,\bar{q}\}$, and fix the
functions in this way
$$\phi(\u)=u,\; f(\u)=p,\;g(\u)=\bar{p},$$
we can write the matrix elements of $R(u,w;p,q;\bar{p},\bar{q})$
as

\bea\label{a_fr} &a_1(u,w;p,q;\bar{p},\bar{q})=
 \frac{p\sinh{(u+u_0)}\sinh{(u_0-w)}}{q (\sinh{u_0})^2}+
  \frac{\bar{q}\sinh{u}\sinh{w}}{\bar{p}(\sinh{u_0})^2},&\\\nn
&a_2(u,w;p,q;\bar{p},\bar{q})=\frac{q\sinh{(w+u_0)}\sinh{(u_0-u)}}{p
(\sinh{u_0})^2}+
  \frac{\bar{p}\sinh{u}\sinh{w}}{\bar{q}(\sinh{u_0})^2},&\\\nn
  &b_1(u,w;p,q;\bar{p},\bar{q})=\frac{\bar{p}\sinh{u}\sinh{(u_0-w)}}{q
  (\sinh{u_0})^2}-
  \frac{\bar{q}\sinh{(u_0-u)}\sinh{w}}{p(\sinh{u_0})^2},&\\\nn
 &b_2(u,w;p,q;\bar{p},\bar{q})=\frac{q\sinh{(w+u_0)}\sinh{u}}{\bar{p}
  (\sinh{u_0})^2}-
  \frac{p\sinh{(u+u_0)}\sinh{w}}{\bar{q}(\sinh{u_0})^2}.&\ena
Taking into account also the parameters followed from the
automorphism  (\ref{auf}), and the notations (\ref{fst}), we can
introduce new parameters $t,\; s$ and write as well  (below now
$\u=\{u,p,\bar{p},t\}$ and $\w=\{w,q,\bar{q},s\}$) the expressions
\bea c_1(\u,\w)=t/s,\quad c_2(\u,\w)=s/t, \ena
for the  $c_i$ elements of the matrix
$R(u,w;p,q;\bar{p},\bar{q};t/s)$.

\subsection{$d_i%(\u,\w)
\neq 0$, $b_i%(\u,\w)
\neq 0$, $i=1,2$}

The case with the choice $d(\u,\w)\neq 0$ brings to the following
 relations for the full functions via the elementary ones (below $\bar{i}=i+1\; \mathrm{mod}(2)$)
\bea\label{abd}
a_i(\u,\w)\!\!\!&=&\!\!\!\frac{d(\u,\w)\Big(d(\u)[a_i(\u)a_i(\w)\!-\!b_{\bar{i}}(\u)b_{\bar{i}}(\w)]+
d(\w)[a_{\bar{i}}(\u)a_{\bar{i}}(\w)\!-\!b_{{i}}(\u)b_{{i}}(\w)]\Big)}{d(\u)^2-d(\w)^2},\\\nn
b_i(\u,\w)\!\!\!&=&\!\!\!\frac{a_{\bar{i}}(\u)b_i(\w)\!-\!a_{\bar{i}}(\w)b_{{i}}(\u)+
d(\u)d(\w)[a_{{i}}(\u)b_{\bar{i}}(\w)\!-\!a_{{i}}(\w)b_{\bar{i}}(\u)]}{d(\u)^2
d(\w)^2-1}.\ena
As previously, we set $f_i(\u,\mathbf{0})=f_i(\u)$, with
$f=a,\;b,\;d$ and $i=1,\;2$ and
 $a_i(\mathbf{0})=1,\; b_i(\mathbf{0})=0,\;
d(\mathbf{0})=0$ following from (\ref{unite}).

Analyzing the YBE with the functions (\ref{abd}) we obtain several
expressions for $d(\u,\w)$ by means of the elementary functions
$f_i(\u),\;f_i(\w)$. Consistency conditions for these expressions
to be equal one to another constitute the relations between the
elementary functions.

Before proceeding farther, let us reveal the nature of the
constants included in the solutions by means of the derivatives of
the elementary functions at the normalization point $\mathbf{0}$.
For the multiparametric
functions $f_i(\w)%=f_i(\{w_1,w_2,...,w_k\})
$ we can suggest that derivation $\frac{d}{d\w}$ is taken along a
path $\{w_i(w_0)\},\; i=1,...,k$,  if the composite
 parameter
 $\w=\{w_1,w_2,...,w_k\}$ is parameterized by a path parameter $w_0$, $w_0 \in \mathcal{C}$.
We imply so, $\frac{d f_i(\w)}{d\w}d\w=(\frac{d
w_1}{dw_0}\partial_{w_1}+ \frac{d w_2}{dw_0}\partial_{w_2}+...+
\frac{d w_k}{dw_0}\partial_{w_k})f_i(\w)dw_0$, and
$f'_i(\mathbf{0})=(\frac{d w_1}{dw_0}\partial_{w_1}+\frac{d
w_2}{dw_0}\partial_{w_2}+...+\frac{d
w_k}{dw_0}\partial_{w_k})f_i(\w)|_{\w=\mathbf{0}}$.  We can see, that the
constant $d_0$ in (\ref{yr1}-\ref{yr3}) can be expressed as
$d_0=-\frac{2 a'(\mathbf{0})}{d'(\mathbf{0})}$. The constants
$\Delta$ and $k$ in (\ref{abc}, \ref{abx}) are equal to
$$ \Delta=\frac{a'_1(\mathbf{0})+a'_2(\mathbf{0})}{b'_1(\mathbf{0})},\quad
k=\frac{b'_2(\mathbf{0})}{b'_1(\mathbf{0})}.$$
It means that the classification of the solutions by means of the
constant parameters of the solutions can be formulated in the
language of the derivatives taken at the normalization point.
Particularly, we have seen for the case $d_i(\u,\w)=0$, that the
solutions with $a'_1(\mathbf{0})+a'_2(\mathbf{0})=0$ describe
free-fermionic models. We shall ascertain that this is a general
property.

So, below, we shall use derivatives at the point $\mathbf{0}$ for
describing the constants.

The  consistency conditions, mentioned at the beginning
of this subsection,  %for different expressions of $d(\u,\w)$ to be
%equivalent,
% for we obtain
 contain the following general relation valid for each case
{ \bea (\star)\quad\mathbf{ 2
d(\u)[a_1'(\mathbf{0})-a_2'(\mathbf{0})]=d'(\mathbf{0})
\Big(a_1(\u)^2-a_2(\u)^2-b_1(\u)^2+b_2(\u)^2\Big)},\label{relat}
\ena}
Let us discuss different cases separately. The next part of this
section is organized as follows. In the first subsection we
discuss the case when the constant in the constraint (\ref{relat})
$(a_1'(\mathbf{0})-a_2'(\mathbf{0}))/d'(\mathbf{0})$ vanishes.
Here we discuss in detail the two possible  situations $b_1(\u)=\pm
b_2(\u)$ ($b_1'(\mathbf{0})=\pm b_2'(\mathbf{0})$), which after
appropriate parametrizations are corresponding respectively to the
solution of the XYZ model and to the
  two-parametric free-fermionic $R$-matrix. In the next
  subsections the solutions with non vanishing constant
   $(a_1'(\mathbf{0})-a_2'(\mathbf{0}))/d'(\mathbf{0})\neq 0$
is presented. It appears that all the  solutions in this case have
the free fermionic property and for them
$(a_1'(\mathbf{0})=-a_2'(\mathbf{0}))$ in general. In the
subsection 2.3.2 we separate two cases with $b_1'(\mathbf{0})=\pm
b_2'(\mathbf{0})$. One solution here can be considered as the
generalization of the free-fermionic XY model's $R$-matrix (for
which $a_i'(\mathbf{0})=0$), the second one corresponds to the
known colored thee-parametric solution \cite{Felder}. In the
subsection 2.3.3 the general free-fermionic solutions are
presented with two arbitrary functions and two non-vanishing
constants.

\subsubsection{$a_1'(\mathbf{0})=a_2'(\mathbf{0})$: $a_1(\u)=a_2(\u)$}

Here we consider the
%case. %we have
 situation with $a_1'(\mathbf{0})=a_2'(\mathbf{0})$. %Considering the
%situation $a_1'(\mathbf{0})= a_2'(\mathbf{0})$, and
%Supposing $a_1(\u)\neq a_2(\u)$, we shall
The analysis of the next equations brings to the
relation $a_1(\u)^2 = a_2(\u)^2$. This means that we must
consider the case $a_1(\u) = a_2(\u)$, as $a_1(\mathbf{0})=
a_2(\mathbf{0})=1$ (\ref{unite}). One can verify, that  the
situation $a_1(\u) = -a_2(\u)$ brings to the rather trivial
solutions.

From the relation (\ref{relat}) we obtain $b_1(\u)=\pm b_2(\u)$.
At first we discuss the case $b_1(\u)=b_2(\u)$.

\paragraph{$\bullet\; b_1(\u)=b_2(\u)$.} This choice immediately implies
 $a_1(\u,\w)=a_2(\u,\w)$,
$b_1(\u,\w)=b_2(\u,\w)$. Let us omit by now the indexes $i=1,2$. So,
in fact, this case is equivalent to the Baxter's discussion of
YBE for the XYZ-model. The functions $f(\u)$ obey to the following
constraints:
\bea \label{dk}&d(\u)=\frac{d'(\mathbf{0})}{b'(\mathbf{0})}a(\u)b(\u),&\\
\label{abdelta}&a(\u)^2+b(\u)^2-1-d(\u)^2=\frac{2a'(\mathbf{0})}{b'(\mathbf{0})}a(\u)b(\u).&\ena
 Similar relations
take place for the functions $f(\u,\w)$ too, with the same
constants $\frac{d'(\mathbf{0})}{b'(\mathbf{0})}\equiv k, \;
\frac{a'(\mathbf{0})}{b'(\mathbf{0})}\equiv \Delta$, as it follows
from the relations (\ref{abd}). Thus we have the following
expressions for the functions $f_i(\u,\w)$ with one arbitrary
function $a(\u)$ and two constants, as the other functions $d(\u)$
and $b(\u)$ can be expressed by means of them using the relations
(\ref{dk}, \ref{abdelta}).
\bea\nn a(\u,\w)&=&\frac{\frac{a(\u)}{a(\w)}(1-b(\w)^2)+(1-k^2
a(\u)^2) b(\u)b(\w)}{1-k^2 a(\u)^2 b(\w)^2},
\\
b(\u,\w)&=&\frac{a(\w)b(\u)-a(\u)b(\w)}{1-k^2
a(\u)a(\w)b(\u)b(\w)},\label{r_xyzuw}\\\nn
 d(\u,\w)&=&k a(\u,\w)b(\u,\w). \ena
%
%As in the previous discussion,
We can set here the usual XYZ elliptic  parameterization, %as the
%constraints give only one arbitrary function $a(\u)$ and two
 %constant parameters $k,\; \Delta$,
placing %passing %to the arbitrary
the arbitrariness of the solution on the argument function
$\phi(\u)$, and the constants $\phi_0$ and $\mathbf{k}$ (elliptic
module), and writing
\bea\label{xyz}
a(\u)=\frac{\mathrm{sn}[\phi(\u)+\phi_0,\mathbf{k}]}{\mathrm{sn}[\phi_0,\mathbf{k}]},\quad
b(\u)=\frac{\mathrm{sn}[\phi(\u),\mathbf{k}]}{\mathrm{sn}[\phi_0,\mathbf{k}]},
\quad \phi(\mathbf{0})=0. \ena
So, here actually %effectively
we have only one-parameteric solution, as the functions
$f_i(\u,\w)$ acquire difference property by this parameterization
$f(\phi(\u),\phi(\w))=f(\phi(\u)-\phi(\w))$.

\paragraph{$\bullet\; b_2(\u)=-b_1(\u)$.}
Now let us turn to the discussion of the case
$b_2(\u)=-b_1(\u)\equiv -b(\u)$. Here %from the series expansion
we %immediately
found that $a'(\mathbf{0})=0$ and
\bea a(\u)^2-b(\u)^2-d(\u)^2-1=0. \label{a-bc} \ena
This is a free-fermionic condition, as it immediately implies
\bea a(\u,\w)^2-b(\u,\w)^2-d(\u,\w)^2-1=0. \ena
Also we have $b_1(\u,\w)=-b_1(\w,\u)=-b_2(\u,\w)$ and
$a_1(\u,\w)=a_1(\w,\u)=a_2(\u,\w)$.
 The relation (\ref{a-bc}) appears to be enough for the
 $\check{R}(\u,\w)$-matrix to satisfy the Yang-Baxter equations.
 So, there are two arbitrary functions in this solution. One
 can choose a parameterization, which will bring to the familiar
 solution (see, e.g. \cite{SSh}). Let $1+d(\u)^2=g(\u)^2$ and
 $a(\u)=g(\u)\cosh{[\phi(\u)]}$, $b(\u)=g(\u)\sinh{[\phi(\u)]}$,
 then the full functions become
\bea \label{raa}a_1(\phi(\u),\phi(\w);g(\u),g(\w))&=&
\frac{\cosh{[\phi(\u)-\phi(\w)]}g(\u)g(\w)}{1+\sqrt{g(\u)^2-1}\sqrt{g(\w)^2-1}},\\
b_1(\phi(\u),\phi(\w);g(\u),g(\w))&=&
\frac{\sinh{[\phi(\u)-\phi(\w)]}
%\Big(\sqrt{g(\u)^2-1}\sqrt{g(\w)^2-1}-1\Big)
g(\u)g(\w)}
{1+\sqrt{g(\u)^2-1}\sqrt{g(\w)^2-1}},\nn\\
d(\phi(\u),\phi(\w);g(\u),g(\w))&=&
\frac{\sqrt{g(\u)^2-1}-\sqrt{g(\w)^2-1}}{1+\sqrt{g(\u)^2-1}\sqrt{g(\w)^2-1}}.\nn
\ena
Setting
\bea g(\u)=\frac{1}{\cos{[\psi(\u)]}}, \label{r_aauw}\ena
 we shall obtain the %known
two-parametric solution  \cite{SSh}
$\check{R}(\phi(\u)-\phi(\w),\psi(\u)-\psi(\w))$ brought in
apparent matrix form in the Appendix (\ref{ybezuv}).

\subsubsection{$a_1'(\mathbf{0})\neq a_2'(\mathbf{0})$%: $a_1(\u)\neq a_2(\u)$.
}
%Now we turn to
For the %more general
case $a_1'(\mathbf{0})\neq a_2'(\mathbf{0})$,
and hence $a_1(\u)\neq a_2(\u)$, %Then
besides of the the relation (\ref{relat})
%
%\be d(\u)=\frac{d'(\mathbf{0})}{2(a_1'(\mathbf{0})-
%a_2'(\mathbf{0}))}\Big(a_1(\u)^2-a_2(\u)^2-b_1(\u)^2+b_2(\u)^2\Big)
%\ee
%
%is true.
 also %Also
the free-fermionic property for the  functions $f_i(\u)$,
$f=a,b,c,d$,  is arisen
\bea
(\star\star)\quad\mathbf{a_1(\u)a_2(\u)+b_1(\u)b_2(\u)=1+d(\u)^2.}\label{fre}\ena

This relation takes place for the functions $f_i(\u,\w)$ too.
Analyzing the next part of the YBE we obtain the following
equation
\be(a_1'(\mathbf{0})+
a_2'(\mathbf{0}))\Big(a_1(\u)b_1(\u)-a_2(\u)b_2(\u)\Big)=0.\label{rep}\ee
 If
to take the relation $a_1(\u)b_1(\u)=a_2(\u)b_2(\u)$ %and hence
%$b_1'(\mathbf{0})=b_2'(\mathbf{0})$,
the next equations give that
\be \mathbf{a_1'(\mathbf{0})=-a_2'(\mathbf{0})}\ee
nevertheless. One could obtain this constraint also directly from
the relation (\ref{fre}) expanding it around the point
$\mathbf{0}$, i.e. it is the peculiarity of the free-fermionic
property providing that the condition (\ref{unite}) takes place. Thus the
relation (\ref{rep}) in some sense is the analog of the relation
(\ref{re}), although all the solutions here are
free-fermionic.  Anyway, let us at first consider the situation with the
constraint $a_1(\u)b_1(\u)=a_2(\u)b_2(\u)$, which is the
feature of the usual eight-vertex model and gives the solution with one arbitrary function and two constants.

\paragraph{$\bullet$ The case $a_1(\u)b_1(\u)=a_2(\u)b_2(\u)$: $b_1'(\mathbf{0})=b_2'(\mathbf{0})$.}

%Note, that
The functions $f_i(\u)$ satisfy the following relations
\bea
&a_1(\u)\left(b_1(\u)^2+a_2(\u)^2\right)=a_2(\u)(1+d(\u)^2),&\label{a1b1}\\
&f_0 b_1(\u)\left(a_1(\u)^2+a_2(\u)^2\right)=2a_2(\u)d(\u),&\label{a2b2}\\
&4 x_f d(\u)
a_2(\u)^2=\left(a_1(\u)^2-a_2(\u)^2\right)\left(a_2(\u)^2+b_1(\u)^2\right),&\label{abxf}\ena
%
%from which can be found the above parametric functions.
The first
relation is just the free-fermionic condition. %for the functions
%$f_i(\u,\mathbf{0})$, $f=a,b,c,d$.
The parameters $x_f$ and $f_0$ are arbitrary constants and are
chosen so, that $x_f=a_1'(\mathbf{0})/d_1'(\mathbf{0})$ and
$f_0=d_1'(\mathbf{0})/b_1'(\mathbf{0})$. Note that for this case
$b_1'(\mathbf{0})=b_2'(\mathbf{0})$.   The solutions can be
parameterized in the following form (below
$x_0=\frac{x_f}{f_0}=\frac{a_1'(\mathbf{0})}{b_1'(\mathbf{0})}$)
\bea &a_1(\u)=\sqrt{1+x_0 f_x(\u)}F_x(\u),\quad
a_2(\u)=\sqrt{1-x_0 f_x(\u)}F_x(\u)&\nn\\
&b_1(\u)=\frac{\sqrt{1-x_0
f_x(\u)}\Big(1-\sqrt{1-f_x(\u)^2}\Big)}{f_x(\u)}F_x(\u),\quad
b_2(\u)=\frac{\sqrt{1+x_0
f_x(\u)}\Big(1-\sqrt{1-f_x(\u)^2}\Big)}{f_x(\u)}F_x(\u)&\nn\\
&d(\u)=\frac{f_0
F_x(\u)^2\Big(1-\sqrt{1-f_x(\u)^2}\Big)}{f_x(\u)},\;\;\;F_x(\u)=
\frac{\sqrt{\sqrt{1-x_0^2
f_x(\u)^2}-\sqrt{1-(x_0^2+f_0^2)f_x(\u)^2}}}{f_0\left(1-\sqrt{1-f_x(\u)^2}\right)}.&\label{a_spec}
\ena
We have introduced the function $f_x(\u)$ as an independent
function which has the property $f_x(\mathbf{0})=0$. Of course it
is possible to choose different parameterizations, taking for
independent function, as example the function $d(\u)$.
 It could seem that here (taking into account the existence of the factors
  with square roots in (\ref{a_spec})) it is reasonable to
impose the elliptic functions, %but let us note that here there are
%two constants.
% But it seems that here there is
%no such "good" parameterization which will bring to the difference
%property for the argument of the $\check{R}$-matrix.
but note, that there are two arbitrary constants in the square
root factors and the elliptic parameterization is significant only when
one of them vanishes. Indeed, when we try to parameterize so, that
the difference property  to be for a
pair of the parameters ($u,w$) from the set $\{\u,\w\}$, $f_i(u,w;...)=f_i(u-w;...)$,
 then we
come to the simple elliptic parameterization with the particular
case $a_1'(\mathbf{0})=-a_2'(\mathbf{0})=0$, i.e. $x_0=0$, and
$a_1(\u)=a_2(\u)$, $b_1(\u)=b_2(\u)$, which is the solution
corresponding to the XY-model, containing one arbitrary constant.

 The complete functions $f(\u,\w)$ can be found from (\ref{abd}), the expression
for $d(\u,\w)$ is brought in the Appendix. Let us
note %at first
that here the condition $a_1(\u)b_1(\u)=a_2(\u)b_2(\u)$ is not
universal, i.e. $a_1(\u,\w)b_1(\u,\w)\neq a_2(\u,\w)b_2(\u,\w)$ at
general. In particular, we have $a_1(\mathbf{0},\u)=a_2(\u)$,
$a_2(\mathbf{0},\u)=a_1(\u)$, $b_1(\mathbf{0},\u)=-b_1(\u)$,
$b_2(\mathbf{0},\u)=-b_2(\u)$ and
$a_1(\mathbf{0},\u)b_2(\mathbf{0},\u)=a_2(\mathbf{0},\u)b_1(\mathbf{0},\u)$.
The free fermionic condition in that context is universal.

One can prove (after some tangled calculations), that the
functions above satisfy to the whole set of YBE, although they can
be obtained only by considering the particular cases of them, when
one of the composite arguments has been taken to be $\mathbf{0}$.
This is a natural result,
 {\textit{as the
point $\mathbf{0}$ is chosen arbitrarily, we could take any point
$\w_0$ for defining the elementary functions
$f_i(\u)=f_i(\u,\w_0)$, and then the primary values $f_i(\w_0)$
for them
 will be fixed from the normalization condition}}
(\ref{unite}). Also the free-fermionic condition can be proved for
the entire functions $f_i(\u,\w)$, $f=a,b,c,d$. Another
parameterization would be given in the Appendix, with a short
description of a receipt how to solve YBE. So, for this particular
case we have the solutions with one arbitrary function and two
arbitrary constants, and in general this is a two-parametric
solution. % (and becomes one-parametric only when one of the
%arbitrary constants vanishes, and $R$-matrix can be parameterized
%so, that it depends on the difference of the parameters).
 $\bullet$

The solution discussed above is the special case of
 the general solutions with the property $a_1'(\mathbf{0})=-
a_2'(\mathbf{0})$.  And in general from the overall
analysis of the YBE we come to the next important constraint %relation
\be (\star\star\star)\quad
\mathbf{d(\u)(b_1'(\mathbf{0})+b_2'(\mathbf{0}))=2
d'(0)(a_1(\u)b_2(\u)+a_2(\u)b_1(\u))}. \label{s_spec}\ee
 This relation together with (\ref{relat}) and
(\ref{fre}) is enough for $f_i(\u)$ to satisfy the whole set of
YBE. The corresponding general solutions will be discussed
in the subsection 2.3.3.
 Here we should like to investigate separately the case with the
vanishing constant $(b_1'(\mathbf{0})+b_2'(\mathbf{0}))/d'(0)=0$
(i.e. $a_1(\u)b_2(\u)+a_2(\u)b_1(\u)=0$) (\ref{s_spec}) which as
we shall see, corresponds to the elliptic three-parametric
solution \cite{BS}.

\paragraph{$\bullet$ The case $a_1(\u)b_2(\u)+a_2(\u)b_1(\u)=0$: $b_1'(\mathbf{0})=-b_2'(\mathbf{0})$.} There are three relations  on the five functions $a_1(\u),\;
a_2(\u),\; b_1(\u),\; b_2(\u),\; d(\u)$, and hence the solutions
contain two arbitrary functions (four-parametric solution) and one
arbitrary constant ($x_f=
\frac{a_1'(\mathbf{0})}{d'(\mathbf{0})}=\frac{a_1'(\mathbf{0})-
a_2'(\mathbf{0})}{2 d'(\mathbf{0})}$, which comes from
(\ref{relat})). From the study of the relations it appears that this is
just the "colored" solution presented in the works
\cite{Felder,BVS}, provided that we require difference property
for one of the pairs of the parameters. Let us describe this
solution.

 Fixing the arbitrary functions as $d(\u)$ and
 $f_z(\u)=b_2(\u)/a_2(\u)$ we find for the remaining functions (below
two auxiliary functions $g_x(\u)$ and $d_f(\u)$ are introduced and
instead of $d(\u)$ the function $d_f(\u)$  could be considered as
elementary function)
\bea
&a_1(\u)=g_x(\u)\sqrt{\frac{1+d(\u)^2}{g_x(\u)(1-f_z(\u)^2)}},\quad
b_1(\u)=
-g_x(\u)f_z(\u)\sqrt{\frac{1+d(\u)^2}{g_x(\u)(1-f_z(\u)^2)}},&\nn\\
&a_2(\u)=\sqrt{\frac{1+d(\u)^2}{g_x(\u)(1-f_z(\u)^2)}},&\label{r_sp}
\\&g_x(\u)= x_f d_f(\u)\pm\sqrt{1+x_f^2 d_f(\u)^2},\quad
d_f(\u)=\frac{2 d(\u)}{1+d(\u)^2}.&\nn \ena
The matrix elements now look like as
\bea
&d(\u,\w)=2\frac{[d(\u)^2-d(\w)^2]}{[1+d(\u)^2][1+d(\w)^2]\Big(d_f(\u)g_x(\w)
+d_f(\w)g_x(\u)-2 x_f d_f(\u)d_f(\w)\Big)},&\\&
a_1(\u,\w)=\frac{2\Big[[1+d(\u)^2][1+d(\w)^2][1-f_z(\u)^2][1-f_z(\w)^2]\Big]^{-\frac{1}{2}}}{%\sqrt{%g_x(\u)g_x(\w)
%[1+d(\u)^2][1+d(\w)^2][1-f_z(\u)^2][1-f_z(\w)^2]}
\Big[d_f(\u)g_x(\w) +d_f(\w)g_x(\u)-2 x_f
d_f(\u)d_f(\w)\Big]}\times &\\&
\nn %\frac{%\Big(d(\u)[g_x(\u)g_x(\w)-f_z(\u)f_z(\w)]+
%d(\w)[1-f_z(\u)f_z(\w)g_x(\u)g_x(\w)]\Big)
\Big(d(\u)\sqrt{g_x(\u)g_x(\w)}+d(\w)\frac{1}{\sqrt{g_x(\u)g_x(\w)}}-f_z(\u)f_z(\w)\Big[
\frac{d(\u)}{\sqrt{g_x(\u)g_x(\w)}}+d(\w)\sqrt{g_x(\u)g_x(\w)}\Big]
%d(\w)[1-f_z(\u)f_z(\w)g_x(\u)g_x(\w)]
%\Big)
\Big)%}
%{\Big(d_f(\u)g_x(\w)
%+d_f(\w)g_x(\u)-2 x_f d_f(\u)d_f(\w)\Big)}
,&\\&
b_1(\u,\w)=\sqrt{\frac{[1+d(\u)^2][1+d(\w)^2]}{%\sqrt{%g_x(\u)g_x(\w)
[1-f_z(\u)^2][1-f_z(\w)^2][-1+d(\u)^2 d(\w)^2]}} \times&\\&\nn
%\frac{g_x(\u)f_z(\u)-g_x(\w)f_z(\w)+d(\u)d(\w)\Big(g_x(\u)f_z(\w)-
%g_x(\w)f_z(\u)\Big)}{-1+d(\u)^2 d(\w)^2}
\left(f_z(\u)\Big[\sqrt{\frac{g_x(\u)}{g_x(\w)}}-d(\u)d(\w)\sqrt{\frac{g_x(\w)}{g_x(\u)}}\Big]
-f_z(\w)\Big[\sqrt{\frac{g_x(\w)}{g_x(\u)}}-d(\u)d(\w)\sqrt{\frac{g_x(\u)}{g_x(\w)}}\Big]
\right),&\\&
b_2(\u,\w)=\sqrt{\frac{[1+d(\u)^2][1+d(\w)^2]}{%\sqrt{%g_x(\u)g_x(\w)
[1-f_z(\u)^2][1-f_z(\w)^2][-1+d(\u)^2 d(\w)^2]}} \times&\\&\nn
%\frac{g_x(\u)f_z(\w)-
%g_x(\w)f_z(\u)+d(\u)d(\w)\Big(g_x(\u)f_z(\u)-g_x(\w)f_z(\w)\Big)}{-1+d(\u)^2
%d(\w)^2}
\left(f_z(\w)\Big[\sqrt{\frac{g_x(\u)}{g_x(\w)}}-d(\u)d(\w)\sqrt{\frac{g_x(\w)}{g_x(\u)}}\Big]
-f_z(\u)\Big[\sqrt{\frac{g_x(\w)}{g_x(\u)}}-d(\u)d(\w)\sqrt{\frac{g_x(\u)}{g_x(\w)}}\Big]
\right),&\\& a_2(\u,\w)=a_1(\w,\u).&\ena
The function $d(\u,\w)$ depends only on the elementary functions
$d_f(\u)$ and $d_f(\w)$. If to demand that $d(\u,\w)=d(\u-\w)$
then we shall have the following differential equation on the
function $d_f(\u)$
\bea d_f'(\u)=d_f'(\mathbf{0})\sqrt{1-d_f(\u)^2}\sqrt{1+x_f^2
d_f(\u)^2}, \ena
the solution of which is  the Jacobi's elliptic function
$\mathrm{sn}[\u,k]$ with $k^2=-x_f^2$. This corresponds to the
colored parameterization presented in \cite{BVS}, if  to take %the choice of
 the composite parameters in the matrix $R(\u,\w)$ to be
$\u=\{u,p\},\;\;\; \w=\{v,q\}$, and and to fix the next arbitrary
functions as $f_z(\u)=p,\;\;\;f_z(\w)=q$. % $x_f
%d_f(\u)=\mathrm{sn}[u,k]$, with $k=i x_f$,
Thus one can check that the whole matrix has difference property
in respect of the
variables $u$ and $w$ - $R(\u,\w)=R(u,w,p,q)=R(u-w,p,q)$. %, as the
%elliptic function parameterization is followed from that
%requirement.
 The full matrix elements are given in the Appendix.
The case with $x_f=0$ has been discussed in the subsection 2.3.1.
$\bullet$

 \subsubsection{The general free-fermionic
  case with $a'(\mathbf{0})=-a_2'(\mathbf{0})$%,\; $b_1'(\mathbf{0})\neq b_2'(\mathbf{0})$
.}

And in general case, at the condition
 $a'(\mathbf{0})=-a_2'(\mathbf{0})$, %when $b_1'(\mathbf{0})\neq \pm b_2'(\mathbf{0})$,
 the solution contains two arbitrary constants and two arbitrary functions
and corresponds to the four parametric solution. The equations on
the elementary functions are the
  already presented three relations ($\star, \star\star, \star\star\star$) (\ref{relat}, \ref{fre}, \ref{s_spec}), which are collected
together below (the constants are denoted as before $x_f
=\frac{a_1'(\mathbf{0})-a_2'(\mathbf{0})}{2d'(\mathbf{0})}$,
$f_0=\frac{2d'(\mathbf{0})}{b_1'(\mathbf{0})+b_2'(\mathbf{0})}$)%following three
\bea %2 d(\u)[a_1'(\mathbf{0})-a_2'(\mathbf{0})]=d'(\mathbf{0})
4 x_f d(\u)=\Big(a_1(\u)^2-a_2(\u)^2-b_1(\u)^2+b_2(\u)^2\Big),\label{relatx}\\
%d(\u)(b_1'(\mathbf{0})+b_2'(\mathbf{0}))=d'(\mathbf{0})
2  d(\u)=f_0
\Big(a_2(\u)b_1(\u)+a_1(\u)b_2(\u)\Big),\label{relatz}\\
a_1(\u)a_2(\u)+b_1(\u)b_2(\u)-1-d(\u)^2=0.\label{relatw} \ena

 The function $d(\u,\w)$ looks similar to the expression of the previous case, only
 the constant parameter $x_f^2$ is changed into $x_f^2-f_0^{-2}$, where
$f_0^{-1}=\bar{x}_0$
\bea d(\u,\w)=\frac{\pm
2[d(\u)^2-d(\w)^2]}{[1+d(\u)^2][1+d(\w)^2]\Big(d_f(\u)\sqrt{1+
(x_f^2-\bar{x}_0^2) d_f(\w)^2} +d_f(\w)\sqrt{1+(x_f^2-\bar{x}_0^2)
d_f(\u)^2}\Big)}. \ena
This means
that the constraint $d(\u,\w)=d(\u-\w)$ %then we shall have the
%following differential equation on the function $d_f(\u)$
%
%\bea
%d_f'(\u)=d_f'(\mathbf{0})\sqrt{1-d_f(\u)^2}\sqrt{1-(x_0^2-x_f^2)d_f(\u)^2},
%\ena
%
%which corresponds
leads to the Jacobi's elliptic function $\mathrm{sn}[\u,k]$ with
$k^2=\bar{x}_0^2-x_f^2$. If to request also that the remained
functions have the difference property, we shall come to a
particular case, with the following differential relation for the
next arbitrary function chosen as $f_g(\u)=b_1(\u)/a_1(\u)$ -
%
%\bea
$  %\\
f_g'(\u)=f_g'(\mathbf{0})\frac{\sqrt{1-d_f(\u)^2}(1-f_g(\u)^2)}{x_f
d_f(\u)+\sqrt{1-(\bar{x}_0^2-x_f^2)d_f(\u)^2}}, \quad
f_g'(\mathbf{0})=1/2 \bar{x}_0 d_f'(\mathbf{0}).
$ %\ena
  When $\bar{x}_0\neq 0$ ($b_1'(\mathbf{0})+b_2'(\mathbf{0})\neq
0$), then we shall have the non-homogeneous free fermionic
solution of the paper \cite{SSh} (corresponding to the 2d Ising
model, the matrix form of which is presented in the Appendix, at
the particular limit, \cite{ShAS}), with the requirement $b_1=b_2$
and the parameters $k^2=\bar{x}_0^2-x_f^2$ and
$\mathrm{dn}[u_0,k]=x_f/\bar{x}_0$.

At the general case the solution can be represented by the
following matrix elements (the expression of the element
$d(\u,\w)$ is written above)
\bea a_1(\u,\w)=\frac{2\sqrt{[1-\bar{x}_0
f_g(\u)d_f(\u)][1-\bar{x}_0 f_g(\w)d_f(\w)]}}{\sqrt{g_x(\u)
g_x(\w)[1+d(\u)^2][1+d(\w)^2][1-f_g(\u)^2][1-f_g(\w)^2]}}\times \\
\nn \frac{\Big(d(\u)[g_x(\u)g_x(\w)-f_p(\u)f_p(\w)]+
d(\w)[1-f_g(\u)f_g(\w)g_x(\u)g_x(\w)]\Big)}{\Big(d_f(\u)G_x(\w)
+d_f(\w)G_x(\u)-2 x_f d_f(\u)d_f(\w)\Big)},\\
b_1(\u,\w)=a_2(\u)a_2(\w)%\sqrt{\frac{[1+d(\u)^2][1+d(\w)^2][1-\bar{x}_0
%f_g(\u)d_f(\u)][1-\bar{x}_0 f_g(\w)d_f(\w)]}{g_x(\u)
%g_x(\w)[1-f_g(\u)^2][1-f_g(\w)^2]}}\times\\\nn
\frac{g_x(\w)f_g(\w)-g_x(\u)f_g(\u)\!+\!d(\u)d(\w)\Big[g_x(\u)f_p(\w)-
g_x(\w)f_p(\u)\Big]}{-1+d(\u)^2 d(\w)^2},\\
b_2(\u,\w)=a_2(\u)a_2(\w)%\sqrt{\frac{[1+d(\u)^2][1+d(\w)^2][1-\bar{x}_0
%f_g(\u)d_f(\u)][1-\bar{x}_0 f_g(\w)d_f(\w)]}{g_x(\u)
%g_x(\w)[1-f_g(\u)^2][1-f_g(\w)^2]}}
%\times\\\nn
\frac{g_x(\u)f_p(\w)-
g_x(\w)f_p(\u)\!+\!d(\u)d(\w)\Big[g_x(\w)f_g(\w)-g_x(\u)f_g(\u)\Big]}{-1+d(\u)^2
d(\w)^2},\\
a_2(\u,\w)=a_1(\w,\u).\label{abcd_gener}\ena
 Here we have two arbitrary and independent functions $d(\u)$ and
 $f_g(\u)$ and two constant $\bar{x}_0$ and $x_f$, the other functions
 are expressed by them through the relations
\bea g_x(\u)=\frac{G_x(\u)}{1-\bar{x}_0 f_g(\u) d_f(\u)},\quad
G_x(\u)=x_f d_f(\u)
\pm \sqrt{1-(\bar{x}_0^2-x_f^2)d_f(\u)^2},\\
f_p(\u)=\frac{-\bar{x}_0 d_f(\u)+f_g(\u)}{-1+\bar{x}_0 f_g(\u)
d_f(\u)},\quad
d_f(\u)=\frac{2 d(\u)}{1+d(\u)^2},\\
a_2(\u)=\sqrt{\frac{[1+d(\u)^2][1-\bar{x}_0 f_g(\u)
d_f(\u)]}{[1-f_g(\u)^2]g_x(\u)}}. \ena
In general we can write the $R$-matrix as
\bea R_{ij}(d(\u),d(\w);f_g(\u),f_g(\w))=R_{ij}(u,w;p,q),
\label{r_fge}\ena
where there are done the following notations
\bea f_g(\u)=p,\quad f_g(\w)=q, \ena
and we have used instead of the composite arguments the usual
spectral parameters $\u=\{u,p\}$, with $d(\u)=d(u)$, $f_g(\u)=p$.

The $sl_q(2)$-invariant solution in  the paper \cite{shkh}
containing one arbitrary function $h(u,\varepsilon)$, two parameters
$\varepsilon_{1,2}$ (the characteristics of the cyclic irreps of
the $sl_q(2)$ algebra at $q=i$) and two arbitrary constants (see
the subsection 4.4 in the mentioned work) belongs to the discussed
here case (\ref{r_fge}) -  the functions $f_g(\u),\; d(\u)$ are
 dependent from the functions $h(u,\varepsilon)$ and $e^{\varepsilon}$ in a specific
nonlinear way (it can be obtained just by comparing the matrix
elements $R_{ij}^{kr}$), which is given in the Appendix.
%The solutions
%$R(\varepsilon_i,\varepsilon_j,\mathbf{x}^a_i,\mathbf{x}^a_j)$ in
%\cite{shkh} are two-parametric in fact, as the dependence from the
%parameters $\mathbf{x}^a_i,\; \mathbf{x}^a_j$, as it was stated
%therein, can be eliminated by the automorphisms (or the basis
%change described in the first section).

%\section{Summary - Conclusions and Outlook}

\begin{table}[t]
{%\footnotesize { }
\scriptsize
%$\;--------------------------------------------$\\
{ }$
========================
========================
===================$\\
 {\textbf{The $4\times 4$ solutions to YBE with eight non-zero matrix
 elements, with primary condition  $R(\u,\u)\!=\!I$ %can be
classified
% (according to the analysis in this paper as)
by means of the
constant  parameters $f_i'(\mathbf{0})$. %into two big groups and
%then
%into three subgroups.
Three columns are overlapped at $f_i'(\mathbf{0})=0$, $f=a,b$.
\\
The subsequent particular cases and the
special cases of the solutions with difference property.\\ %are also presented.
 The constants are denoted as  $x_f=\frac{a_1'(\mathbf{0})}{d'(\mathbf{0})}$,
  $\bar{x}_0=\frac{b_1'(\mathbf{0})+b_2'(\mathbf{0})}{2 d'(\mathbf{0})}$,
$x_0=\frac{x_f}{\bar{x}_0}$
 }}
\\
{
}$\;\;\;\;\;\;\;\;-----------------------------------------------------$
\\
 \begin{tabular}[h]{ccc}
 $a_1'(\mathbf{0})= a_2'(\mathbf{0})$ &{}& $a_1'(\mathbf{0})=-
 a_2'(\mathbf{0})$\\
 {\begin{tabular}[h]{cc}%$\swarrow$&$\searrow$
\\
 $a_1(\u)=a_2(\u)\;\&\;b_1(\u)=b_2(\u)$&$a_1(\u)=a_2(\u)\;\&\;b_1(\u)=\!\!-b_2(\u)$\\
 generalized $XYZ$-type solutions &free-fermionic solutions\\
  with one arbitrary function&with two arbitrary functions\\
  and two constants:\\
%  $x_0=\frac{a_1'(\mathbf{0})}{d'(\mathbf{0})}$,
%  $x_z=\frac{b_1'(\mathbf{0})+b_2'(\mathbf{0})}{d'(\mathbf{0})}$
 \end{tabular}}&{}&{\begin{tabular}[h]{c}%$\downarrow$
\\$a_1(\u)\neq a_2(\u)$\\
 free-fermionic solutions
 \\ with two arbitrary functions\\
 and two constants $R(p,q,u,w)$ \end{tabular}}\\{ }
  { }{\begin{tabular}[h]{c} \\{\it{
  %the consistency relations
  the complete set of the independent equations on the functions $f_1(\u)$}}
 \end{tabular}} &{ }&{\begin{tabular}[h]{c}\\{\it{the complete set of the independent %consistency conditions
 }}\end{tabular}}\\
{\begin{tabular}[h]{cc} %{\it{the consistency relations }} &{" "}\\
%{ \it on the functions $f_1(\u)$}&{}\\
 %$\downarrow$&$\downarrow$\\
$\frac{a_1(\u)^2+b_1(\u)^2-1-d(\u)^2}{2
a_1(\u)b_1(\u)}=%\frac{a_1'(\mathbf{0})}{b_1'(\mathbf{0})}
x_0 $ &
$\;a_1(\u)^2-b_1(\u)^2=1+d(\u)^2$\\
$\bar{x}_0 d(\u)=%\frac{d'(\mathbf{0})}{b'(\mathbf{0})}x_z
a_1(\u)b_1(\u)$&{ }
 \end{tabular}} &{ }&{\begin{tabular}[h]{c}%\\{\it{the consistency conditions }}\\
 {\it{  equations on the functions $f_i(\u)$}}
 \\$a_1(\u)a_2(\u)+b_1(\u)b_2(\u)=1+d(\u)^2$\\
{$d(\u)=\frac{a_1(\u)^2-a_2(\u)^2-b_1(\u)^2+b_2(\u)^2}{4 x_f}$}\\
$d(\u)=\frac{a_1(\u)b_2(u)+a_2(\u)b_1(\u)}{\bar{x}_0}$ \end{tabular}}\\
 { }{\begin{tabular}[h]{c} {\it{the special cases of the solutions with difference property}}
 \end{tabular}} &{ }&{\begin{tabular}[h]{c}\\{\it{ particular cases }}\end{tabular}}\\
{\begin{tabular}[h]{cc}% {\it{the special case of the solutions }} &{" "}\\
%{ \it with difference property}&{}\\
 %$\downarrow$&$\downarrow$\\
$R_{XYZ}(u-w)$  & $\qquad
R(u_1-w_1,u_2-w_2)$ \\
(\ref{xyz}, \ref{ybezi}) & $\qquad$ (\ref{r_aauw}, \ref{ybezuv})
 \end{tabular}} &{ }&{\begin{tabular}[h]{cc}%\\{\it{ particular cases}}&{\it " "}\\
% $|$&$\downarrow$\\
 %$|$&$b_1'(\mathbf{0})=b_2'(\mathbf{0})$\\$\downarrow$&(\ref{a_spec})\\
 %$b_1'(\mathbf{0})=-b_2'(\mathbf{0})$&{$R(u,w)$}\\
 %(\ref{s_spec}),(\ref{r_sp})&{} \\
% $R(p,q,u-w)$ &{}\end{tabular}}
 $b_1'(\mathbf{0})=-b_2'(\mathbf{0})$&{$b_1'(\mathbf{0})=b_2'(\mathbf{0})$}\\
  (\ref{r_sp}, \ref{color})&(\ref{a_spec})\end{tabular}}
 \end{tabular}
}
\\{%\textcolor[rgb]{0.98,0.00,0.00}{
%{ $\asymp\asymp\asymp\asymp\asymp\asymp\asymp\asymp\asymp\asymp
%\asymp\asymp\asymp\asymp\asymp\asymp\asymp\asymp\asymp\asymp
%\asymp\asymp\asymp\asymp\asymp\asymp\asymp\asymp\asymp\asymp
%\asymp\asymp\asymp\asymp\asymp\asymp\asymp\asymp\asymp\asymp
%\asymp\asymp\asymp\asymp\asymp\asymp\asymp\asymp\asymp\asymp $}%}
%\\
{\footnotesize \scriptsize$
=======================
=======================
=======================
$}
%\\
%
{\footnotesize\scriptsize
%$\vee\wedge\vee\wedge\vee\wedge\vee\wedge\vee\wedge\vee
%\wedge\vee\wedge\vee\wedge\vee\wedge\vee\wedge\vee
%\wedge\vee\wedge\vee\wedge\vee\wedge\vee\wedge\vee
%\wedge\vee\wedge\vee\wedge\vee\wedge\vee\wedge\vee
%\wedge\vee\wedge\vee\wedge\vee\wedge\vee\wedge\vee$
}
}%\\
 {\scriptsize\textbf{\centering{The  case with six non-zero matrix
 elements: $d(\u,\w)=0$} %, which can have in general six parameters, two spectral ones
 %and two pairs of colored ones
}\\
{
}$\;\;\;\;\;\;\;\;-----------------------------------------------------$
\\
 \begin{tabular}[h]{cc}
 $a_1'(\mathbf{0})= a_2'(\mathbf{0})\qquad$&$\qquad a_1'(\mathbf{0})=- a_2'(\mathbf{0})$\\
\textit{independent equations on the functions} $f_i(\u)$ &
\textit{independent equations on the functions} $f_i(\u)$ \\
 (\ref{ab_abc})& (\ref{a12})\\generalized $XXZ$-type solutions
  with two arbitrary functions& free-fermionic solution with three arbitrary functions:\\
  and two constants:&\\
 $R(u-w;p,q)$ (\ref{rpqst})& $R(u,w;p,q,\bar{p},\bar{q})$ (\ref{a_fr})
%  $x_0=\frac{a_1'(\mathbf{0})}{d'(\mathbf{0})}$,
%  $x_z=\frac{b_1'(\mathbf{0})+b_2'(\mathbf{0})}{d'(\mathbf{0})}$
 \end{tabular}}

 \caption{Classification of the $R_{22}$-solutions}\label{tab1}
 \end{table}

\section{Main Conclusions and Outlook%: Statements
} \addtocounter{section}{0}\setcounter{equation}{0}

 We can summarize the results and conclusions following from the
analysis performed in %this paper
the Section 2 regarding the $4\times 4$ solutions of the YBE in
the following interrelated points (statements).

\paragraph{$\bullet$} Yang-Baxter equations define the number of
the arbitrary colored parameters (independent functions), on which
the $R$-matrices can be dependent. For the $XYZ$-type $4\times 4$
multi-parametric $R$-matrices, the maximal number of the relevant
parameters is six (or eight, if we take into account also the
parameters connected with the automorphisms (\ref{auf})).

\paragraph{$\bullet$} As it was said in the Introduction
the scattering matrices of the relativistic particles in $1+1$
theory depends on the difference of the rapidities and the YB
equations, which in this case constitute the factorization
behavior of the multi-particle scattering matrices
\cite{Z,Z1,grs}, actually  depend only on two spectral
parameters. We have shown, that the YBE has the mentioned
"relativity" property, even if/when the matrices $R(\u,\w)\neq
R(\u-\w)$, in the following sense. When we obtain the solutions to
YBE (\ref{ybe}) for the fixed values of one of the composite
parameters, e.g. $\w=\mathbf{0}$, with defined initial
conditions followed from (\ref{unite}), the solution is valid also
for arbitrary $\w$.

\paragraph{$\bullet$} Starting from the usual YBE with two
parametric $R(u,w)$ we can obtain all the multi-parametric
solutions, as when in the solution we get the arbitrary, non-fixed
function, we can regard it as another arbitrary parameter.

\paragraph{$\bullet$} The difference property of the dependence
of the $R$-matrices from the spectral parameters is natural
property in the scattering theory, when the matrix plays the role
of the scattering matrix of two particles, and the spectral
parameters are simply the rapidities ($u,\; w$). When the
particles have additional extra symmetries (are "colored"), the $R$-matrix
in respect of the
parameters, which describe their extra characteristics ("colored"
parameters $p,\;q$) may have no difference property, as it was
in the well known solutions \cite{Felder,BVS,RKS}. The performed
analysis shows, that the number of the independent extra
characteristics, which can be shown in the $R$-matrix, is
restricted (maximally two kind of "colored" parameters, if we do
not take into account the automorphism considered in the first
part of the paper), conditioned by the number of the arbitrary
functions in the YBE solutions. In the cases, when we have no
difference property for any pair of the parameters, but there are some
special values of the constants, at which for a pair of the
parameters the difference property can be recovered, we may consider
again such parameters as "rapidities".

\paragraph{$\bullet$} Now let us turn to the constants existing
in the solutions.
We see, that there can be utmost two arbitrary constants, which
 are arisen naturally in the solutions. In the discussions they
 are
noted  by $k,\;\Delta$ or $x_0,\; f_0$  (we are neglecting the
constants, which can be introduced via the arbitrary
functions, e.g. as in (\ref{a_fr}), and which %do not show
%themselves
%are not appeared in the solutions manifestly and
do not play any role in the classification of the solutions). When
all the eight matrix elements are non-zero, then the
classification of the solutions leads to definite relations on the
elementary functions and constants formulated by means of the
derivatives given at the normalization point. A principle one is
$a_1'(\mathbf{0})=\pm a_2'(\mathbf{0})$. For the "plus" sign, this
relation implies $a_1(\u)=a_2(\u)$ and $b_1(\u)=\pm b_2(\u)$. For
the "minus" sign we have $a_1(\u)\neq a_2(\u)$ and free-fermionic
conditions. The algebraic equations on the functions $f_i(\u)$,
$f=a,b,d$, $i=1,2$ for both cases are three and are dependent on
the constants $(b_1'(\mathbf{0})+b_2'(\mathbf{0}))/d'(\mathbf{0})$
and
$a_1'(\mathbf{0})/d'(\mathbf{0})$. %So, by means of these constants
%$f'_i(\mathbf{0})$ we can classify the solutions.

In the
%following
table Tab.\ref{tab1} we collect together the obtained results on
the principal types of the $4\times 4$ solutions. The cases with
eight and six non-zero entries are considered separately, as just
by taking $d(\u)\to 0$ we could
 not recover all the multi-parametric solutions.

{\textit{Note}}, that one can choose parameterizations so, that
$f'_i(\mathbf{0})=0$ for all the functions $f_i$, but the
constants do not vanish, as they must not depend on the
parameterization, and then one can use instead the higher
derivatives at that point. From this observation it could seem,
that the classification in terms of the derivatives of the
functions at the given point is dependent from the
parameterization, but as
%Anyway,
 the relations, which we use for classification for the solutions %by means
%of the first derivatives
are correct relations for {\it{any and each}} parameterization,
this classification is justified and must be understood in that
sense.
\subsection{ General statements % and $R_{33}$-solution
}

 In usual, solving the system of YBE we take into
account, that in respect  of the functions $f_i$ with any of the
arguments ($\u,\w$), ($\u,\v$) or ($\v,\w$), the YB equations form
 system of homogeneous linear equations. And it immediately gives
  consistency conditions
 for the solutions following from the
 requirement of vanishing of the determinants of the matrices formed by
 means of the
 appropriate coefficients, as it was for the
 symmetric eight-vertex model \cite{Baxt}.
  For large matrices, with considerable amount of different
 matrix elements, such determinants in general can be not so easily  factorizable
 and hence not so much informative about the
 nature of the solutions. We shall adhere to another way for solving YBE -
  a rather simple algebraic
 way, described below, and implemented in the analysis for the
 case $n=2$ performed in the previous
 sections.

 %We denote the following expressions as
%
%\bea
%&R_{33}(\u,\w)=&\\
%\scriptsize
%\nn & \small\left(\ba{ccccccccc}a_1(\u,\w)&0&0&0&0&0&0&0&0
%\\0&b_1(\u,\w)&0&c_1(\u,\w)&0&0&0&0&0\\
%0&0&\bar{b}_1(\u,\w)&0&0&0&c_3(\u,\w)&0&0\\
%0&c_1(\u,\w)&0&b_2(\u,\w)&0&0&0&0&0\\
%0&0&0&0&a_2(\u,\w)&0&0&0&0\\
%0&0&0&0&0&\bar{b}_2(\u,\w)&0&c_2(\u,\w)&0\\
%0&0&c_3(\u,\w)&0&0&0&b_3(\u,\w)&0&0\\
%0&0&0&0&0&c_2(\u,\w)&0&\bar{b}_3(\u,\w)&0\\
%0&0&0&0&0&0&0&0&a_3(\u,\w) \ea\right)& \ena
%

%We shall explore all the possible cases in detail.

 Let us do the notations
\bea \mathcal{K}_{ijk}[\u,\v,\w]\equiv
R_{ij}(\u,\v)R_{ik}(\u,\w)R_{jk}(\v,\w)-R_{jk}
(\v,\w)R_{ik}(\u,\w)R_{ij}(\u,\v), \ena
YB equations now read as $\mathcal{K}_{ijk}[\u,\v,\w]=0$. We shall
consider the matrices with the property (\ref{unite}).  Let us
recall once again, that the parameters $\u,\;\w$ are synthesized,
joint parameters with meaning of the sets of the parameters
$\{u\},\;\{w\}$, connected with the corresponding states. The
matrix elements at the point $(\u,\mathbf{0})$, where $\mathbf{0}$
symbolically denotes a normalization point, we regard as
elementary functions $f_i(\u,\mathbf{0})=f_i(\u)$. Solving YBE, we
consider the following scheme.

\begin{itemize}
\item We fix  $f_i(\u,\mathbf{0})=f_i(\u)$, with the initial
conditions $f_i(\mathbf{0})$, followed from (\ref{unite}) and
consider the equations
$\mathcal{K}_{ijk}[\u,\w,\mathbf{0}]=0$${}^{\ast}$.

\item We express the two-composite parametric functions in terms
of the elementary functions by solving the equations $({ }^\ast)$
\\ -
$f_i(\u,\w)=\mathcal{F}_i[\{f_j(\u),f_j(\w)\}]$${}^{\ast\ast}$.

\item  We find the all set of the equations put on the elementary
functions  by the YBE (${}^{\ast}$) and also
 ($\mathcal{K}_{ijk}[\u,\mathbf{0},\w]=0,\;\mathcal{K}_{ijk}[\mathbf{0},\u,\w]=0$), after inserting there %in the YBE
the complete functions $\mathcal{F}_i\;({}^{\ast\ast})$ \\ -
$\mathcal{G}_k[\{f_j(\u),f_j(\w)\}]=0$${}^{\ast\ast \ast}$.

\end{itemize}

The number of the independent equations $\mathcal{G}_k=0$ defines
the number of the arbitrary elementary functions which can be in
solutions, or, in another words, the number of the independent
parameters - possible colors. As the point $\mathbf{0}$ is chosen
arbitrarily, the solutions must satisfy also to the complete
equations $\mathcal{K}_{ijk}[\u,\v,\w]$. For the all obtained in
this paper solutions it is verified by direct calculations. The
sets $\mathcal{G}_k$ for the $4 \times 4$-matrices are presented
apparently in the Table \ref{tab1} for each case.

%Let us explore the matrices having the property of "particle
%number" conservation ( by mod(2), i.e. $\mathcal{Z}_2$ grading
%symmetry). In matrix representation it means $ R_{ij}^{kr}\neq 0$
%if $i+j+k+r=0 \; (mod\; 2)$.
%
\section{Colored $9\times 9$  %$R_{33}$-
solutions to YBE}

 The simplest known $9\times 9$
solution is the $sl_2$-invariant solution of the form
$R(u,w)=P+(u-w)I$, $P$ is the permutation operator and $I$ is the
unit matrix \cite{GU,JKL,BS}. We shall consider here the simplest
colored generalization of this solution, with the non-vanishing
elements $R_{ij}^{ij}$ and $R_{ij}^{ji}$,
 where the indexes $i,j...$ take the values $1,\; 2,\;3$.

\bea
&R_{33}(\u,\w)=&\\
%\scriptsize
\nn & \small\left(\ba{ccccccccc}a_1(\u,\w)&0&0&0&0&0&0&0&0
\\0&b_1(\u,\w)&0&\bar{c}_1(\u,\w)&0&0&0&0&0\\
0&0&b_3(\u,\w)&0&0&0&\bar{c}_3(\u,\w)&0&0\\
0&c_1(\u,\w)&0&\bar{b}_1(\u,\w)&0&0&0&0&0\\
0&0&0&0&a_2(\u,\w)&0&0&0&0\\
0&0&0&0&0&b_2(\u,\w)&0&\bar{c}_2(\u,\w)&0\\
0&0&c_3(\u,\w)&0&0&0&\bar{b}_3(\u,\w)&0&0\\
0&0&0&0&0&c_2(\u,\w)&0&\bar{b}_2(\u,\w)&0\\
0&0&0&0&0&0&0&0&a_3(\u,\w) \ea\right)& \ena
%

%At first
We  study the %complete and physically motivated
cases, with $a_i\neq 0,\;c_i\neq 0$ and with the condition
(\ref{unite}).
% (below $\check{R}=P R$, where $P$
%is a permutation operator changing the positions of the states,
%and $\mathbf{I}$ is unit operator)
%
%\bea \check{R}(\u,\u) =\mathbf{I}. \label{unite}\ena
%
 The simplest equations followed from YBE (\ref{ybe}) are
\bea &c_i({\u}_1,\u_2)c_i(\u_2,\u_3) \bar{c}_i(\u_1,\u_3) -
 c_i(\u_1,\u_3) \bar{c}_i(\u_1,\u_2) \bar{c}_i(\u_2,\u_3)=0,&\label{eqc}
 \ena
 for $i=1,2,3$. Then we have the following equations
 \bea
&c_2(\u_2,\u_3) \Big(b_1(\u_1,\u_2) b_3(\u_1,\u_3) -
  b_1(\u_1,\u_3) b_3(\u_1,\u_2)\Big)=0,&\label{eqx1}\\
& c_3(\u_2,\u_3) \Big(\bar{b}_1(\u_1,\u_2) b_2(\u_1,\u_3) -
 \bar{b}_1(\u_1,\u_3) b_2(\u_1,\u_2)\Big)=0,&\label{eqx2}\\
 & c_1(\u_2,\u_3) \Big(\bar{b}_3(\u_1,\u_2) \bar{b}_2(\u_1,\u_3) -
 \bar{b}_3(\u_1,\u_3) \bar{b}_2(\u_1,\u_2)\Big)=0,&\label{eqx3}\ena
with similar equations  when instead of $c_i$ the functions
$\bar{c}_i$ stand. The next simple equations are
\bea &c_1(\u_1,\u_2) \Big(b_3(\u_2,\u_3) b_2(\u_1,\u_3) -
 b_3(\u_1,\u_3)
b_2(\u_2,\u_3)\Big)=0,&\label{eqbc1}\\
& \bar{c}_2(\u_1,\u_2) \Big(\bar{b}_1(\u_2,\u_3)
\bar{b}_3(\u_1,\u_3) - \bar{b}_1(\u_1,\u_3)
\bar{b}_3(\u_2,\u_3)\Big)=0,&\label{eqbc2}\\
 & c_3(\u_1,\u_2) \Big(b_1(\u_2,\u_3)
\bar{b}_2(\u_1,\u_3) -
 b_1(\u_1,\u_3)
\bar{b}_2(\u_2,\u_3)\Big)=0.&\label{eqbc3} \ena

 The general solutions to the above equations we
can parameterize by means of the elementary functions %defined as
$f_i(\u)\equiv f_i(\u,\mathbf{0})$, according to the previous discussions, %where again we have introduced a
%symbolical normalization point $\mathbf{0}$,
and  the constants in the solutions we shall parameterize by the
derivatives of the elementary functions taken at the point
$\u=\mathbf{0}$.  The relations which follow from the above
equations (\ref{eqx1}-\ref{eqbc3}),  supposing that
$R_{ij}^{ji}\neq
 0$, when $i\neq j$, can be
presented as %follows %an arbitrary function $g(\u)$ and an
%arbitrary constant $d_0$
%
\bea
\frac{\bar{c}_i(\u,\w)}{c_i(\u,\w)}=\frac{\bar{c}_i(\u)c_i(\w)}{c_i(\u)\bar{c}_i(\w)},\quad
\frac{\bar{b}_1(\u,\w)}{b_2(\u,\w)}=\frac{\bar{b}_1(\u)}{b_2(\u)},\;
\frac{\bar{b}_2(\u,\w)}{\bar{b}_3(\u,\w)}=\frac{\bar{b}_2(\u)}{\bar{b}_3(\u)},\;
\frac{{b}_3(\u,\w)}{{b}_1(\u,\w)}=\frac{{b}_3(\u)}{{b}_1(\u)},\\
\bar{b}_2(\u)=\frac{\bar{b}_2'(\mathbf{0})}{{b}_1'(\mathbf{0})}{b}_1(\u),\quad
b_3(\u)=\frac{b_3'(\mathbf{0})}{b_2'(\mathbf{0})}b_2(\u),\quad
\bar{b}_3(\u)=\frac{\bar{b}_3'(\mathbf{0})}{\bar{b}_1'(\mathbf{0})}\bar{b}_1(\u).\label{constr}
\ena
%
 % and not
%just single ones.

%As it is known there is a possibility to redefine the matrix
%elements of the matrix $R_{ij}$ by the following transformations
%
%\bea R_{n_i n_j}^{p_i p_j}(\u_i,\u_j)\Rightarrow R_{n_i n_j}^{p_i
%p_j}(\u_i,\u_j)\frac{\mathrm{f}_{n_i}(\u_i)\mathrm{f}_{n_j}(\u_j)}
%{\mathrm{f}_{p_i}(\u_i)\mathrm{f}_{p_j}(\u_j)}, \label{aufx} \ena
%
%induced from the transformations $\mathrm{e}_{n_{i,j}}\to
%\mathrm{f}_{n_{i,j}}\mathrm{e}_{n_{i,j}}$  of the vector basis
%$\mathrm{e}_{n_{i,j}}$ ($n_{i,j}=0,1$, $p_{i,j}=1,2,3$) of the
%space $V_{i,j}$.
Via the the transformations (\ref{auf}),  now with
$n_{i,j}=1,2,3$, $p_{i,j}=1,2,3$, which affect only the elements
$R_{ij}^{ji}$, $i \neq j$, we can make two pairs of the symmetric
matrix elements identical one to another. Taking the functions in
this way
$\left(\frac{\mathrm{f}_{2}(\u)}{\mathrm{f}_{1}(\u)}\right)^2=
\alpha_1 \frac{\bar{c}_1(\u)}{c_1(\u)}$,
$\left(\frac{\mathrm{f}_{3}(\u)}{\mathrm{f}_{1}(\u)}\right)^2=
\alpha_3 \frac{\bar{c}_3(\u)}{c_3(\u)}$ the following elements become equal -
$\bar{c}_i(\u)=c_i(\u)$, $i=1,3$. For $i=2$ after this
choice that equality does not take place in general, as the fraction
$\left(\frac{\mathrm{f}_{2}(\u)}{\mathrm{f}_{3}(\u)}\right)$ is
fixed by the previous relations, and by such basis transformations
we could make $\bar{c}_2=c_2$, only if $c_2\approx c_1 c_3$. So,
we shall consider in the following $\bar{c}_1=c_1$,
$\bar{c}_3=c_3$, $\bar{c}_2\neq c_2$. Of course there is an
arbitrariness in our choice, as we could take to be equal the
other two pairs
 ($i=2,3$ or $i=1,2$) as well and consider the situations
with $\bar{c}_1\neq c_1$ or $\bar{c}_3\neq c_3$.

As any solution to YBE is defined up to a multiplicative function,
we can fix also $a_1(\u,\w)=1$. So, initially we  have $12$
elementary functions $f_i(\u)$. Then the analysis of the whole set
of YBE by means of the scheme defined in the Section 3 shows, that
after definition of the functions $\mathcal{F}_i\; (^{\ast\ast})$,
the remaining equations give nine constraints $\mathcal{G}_i\;
(^{\ast\ast\ast})$
 on the functions $f_i(\u)$, including the three relations
 (\ref{constr}).
 We shall not
present the large expressions of the constraints $\mathcal{G}_i$
and will present immediately  the solutions. So, there are three
independent functions in the solutions. We take for
distinctness the functions $c_i(\u)$ as arbitrary ones. %and then
%express the remaining functions using the equations $({}^{\ast\ast\ast})$.
 The result,
after doing the following successive redefinitions\\
$c_2(\u)\equiv f(\u)c_1(\u)c_3(\u)$, $R_{33}(\u,\w)\to
[f(\u)+x_f(1-f(\w))]R_{33}(\u,\w)$, $c_1(\u)=
\frac{\mathbf{c}_1(\u)\sqrt{x_f+f(\u)(1-x_f)}}{f(\u)}$ and
$c_3(\u)= \frac{\mathbf{c}_3(\u)\sqrt{x_f+f(\u)(1-x_f)}}{f(\u)}$,
can be presented as follows
\bea &%f(\u)=\frac{c_2(\u)}{c_1(\u)c_3(\u)},\quad
a_1(\u,\w)=f(\u)+x_f(1-f(\w)), \label{r33}&\\\nn
&a_2(\u,\w)=\frac{\mathbf{c}_1(\u)^2}{\mathbf{c}_1(\w)^2}
%\frac{(1+f(\u)-f(\w))f(\w)+x_0(f(\u)-1)}{f(\w)+x_0(f(\w)-1)}
\left(\frac{(\alpha+1)[f(\u)+x_f(1-f(\w))]+(1-\alpha)[f(\w)+x_f(1-f(\u))]}{2}\right),&\\\nn&
a_3(\u,\w)=\frac{\mathbf{c}_3(\u)^2}{\mathbf{c}_3(\w)^2}
%\frac{(1+f(\u)-f(\w))f(\w)+z_0(f(\u)-1)}{f(\w)+z_0(f(\w)-1)}
\left(\frac{(\bar{\alpha}+1)[f(\u)+x_f(1-f(\w))]+(1-\bar{\alpha})[f(\w)+x_f(1-f(\u))]}{2}\right),&\\\nn&
 c_1(\u,\w)=\frac{\mathbf{c}_1(\u)
 \sqrt{x_f+f(\u)(1-x_f)}\sqrt{x_f+f(\w)(1-x_f)}}{\mathbf{c}_1(\w)},
 \;c_3(\u,\w)=\frac{\mathbf{c}_3(\u)\sqrt{x_f+f(\u)(1-x_f)}\sqrt{x_f+f(\w)(1-x_f)}}{\mathbf{c}_3(\w)},&\\\nn
 &c_2(\u,\w)=\frac{\mathbf{c}_1(\u)\mathbf{c}_3(\u)[x_f+f(\u)(1-x_f)]}{\mathbf{c}_1(\w)\mathbf{c}_3(\w)},\;
 \bar{c}_2(\u,\w)=\frac{\mathbf{c}_1(\u)\mathbf{c}_3(\u)[x_f+f(\w)(1-x_f)]}{\mathbf{c}_1(\w)\mathbf{c}_3(\w)},&\\\nn
 & b_1(\u,\w)=%\frac{c_1(\u)^2(f(\u)-f(\w))}{\alpha},
 \mathbf{c}_1(\u)^2(f(\u)-f(\w)),\;
 b_3(\u,\w)=%\frac{\varphi \;c_3(\u)^2(f(\u)-f(\w))}{\alpha}
 x_f \mathbf{c}_3(\u)^2(f(\u)-f(\w)),
 &\\\nn &\bar{b}_1(\u,\w)=x_f %\alpha
 \frac{f(\u)-f(\w)}{\mathbf{c}_1(\w)^2},\;
b_2(\u,\w)=x_f %\varphi/\alpha
 \frac{\gamma\;
\mathbf{c}_3(\u)^2[f(\u)-f(\w)]}{\; \mathbf{c}_1(\w)^2},&\\\nn
 &\bar{b}_3(\u,\w)=%\alpha/\ \varphi
 \frac{f(\u)-f(\w)}{\mathbf{c}_3(\w)^2},\;
 \bar{b}_2(\u,\w)=%\alpha\\varphi
 \frac{\mathbf{c}_1(\u)^2[f(\u)-f(\w)]}{  \gamma\; \mathbf{c}_3(\w)^2}.&
 \ena
%
%The constants $\alpha,\; \varphi\; \gamma$ are arbitrary, whereas
The numbers %$x_0$ and $z_0$ can take only the values $0$ and $-2$
$\alpha,\; \bar{\alpha}$ can have %take
 only the values $1$ and
$-1$. When the constant
 $x_f=1$ (i.e $\bar{c}_2(\u,\w)=c_2(\u,\w)$), then with respect to the function
$f(\u)$ the solution acquires the difference property
$R_{33}(f(\u),f(\w);\mathbf{c}_i(\u),\mathbf{c}_i(\w))=
R_{33}(f(\u)-f(\w);\mathbf{c}_i(\u),\mathbf{c}_i(\w))$. If
to take %$\alpha=1,\; \varphi=1\; \gamma=1$ and $x_0=z_0=0$,
$x_f=1$, $\gamma=1$, $\alpha=\bar{\alpha}=1$ and
$\mathbf{c}_1(\u)=\mathbf{c}_3(\u)=1$, $f(\u)=1+u$, we shall
recover the %first
known %isotropic %$sl_2$-invariant
$R_{33}$-matrix, which has the form $I+(u-w)P$ \cite{GU,JKL,BS}.
And similarly, at the same particular homogeneous   case ($x_f=1$)
the other rational limits of the solutions (\ref{r33}) can
coincide with the corresponding rational solutions
 in
%In
 the recent paper \cite{33}, where  the authors considered matrices
with more non-vanishing elements, constrained
with  definite symmetry relations on the matrix elements. %The
%constants $\alpha,\; \varphi$ we can hide in the functions
%$c_i(\u)$, so
 The  $N=3$ case
 of the general $U(1)^{N-1}$-symmetric $R$-matrices, discussed in
 \cite{33osp}, %it seems,
 can coincide with the matrix (\ref{r33}) after some symmetry
 transformations, as one of the authors of \cite{33osp}
  has kindly checked \footnote{R. Pimenta, e-mail correspondence.}. If to
 require that difference property takes place then the matrix
 (\ref{r33}) would be equivalent to the Perk-Schultz solution for
 the
 three-state case \cite{PerkSch,PerkAu-Yang} (we are thankful to prof. J. Perk for drawing our attention
  to this point), where
 the function $f(u)$ is fixed by  trigonometric function
 $(e^u-x_f)/(1-x_f)$. Then setting $x_f=e^{\eta}$, and with
  appropriate choice of the remaining functions and constants, the
  solution in (21) of the encyclopedia article \cite{PerkAu-Yang}
  can be recovered. The
   case $x_f=1$ corresponds to the rational limit.

Here the general solution has actually one important arbitrary
constant -
$x_f=\frac{c_1'(\mathbf{0})-\bar{c}_2'(\mathbf{0})+c_3'(\mathbf{0})}
{c_1'(\mathbf{0})-c_2'(\mathbf{0})+c_3'(\mathbf{0})}$, one
rapidity parameter, the role of which plays the function $f(\u)$,
and two type of colors, described by functions
$p=\mathbf{c}_1(\u),\; q=\mathbf{c}_1(\w)$ and
$\bar{p}=\mathbf{c}_3(\u),\; \bar{q}=\mathbf{c}_3(\w)$. We can
denote
 the obtained matrices as
 \be
R^{\alpha\bar{\alpha}}_{33}(u,w;p,q;\bar{p},\bar{q}), \qquad
\alpha,\bar{\alpha}=\pm 1. \ee
 Also
 the
 irrelevant colors $s=\frac{\mathrm{f}_2(\u)}{\mathrm{f}_1(\u)}$,
$\bar{s}=\frac{\mathrm{f}_2(\w)}{\mathrm{f}_1(\w)}$,
$t=\frac{\mathrm{f}_3(\u)}{\mathrm{f}_1(\u)}$ and
$\bar{t}=\frac{\mathrm{f}_3(\w)}{\mathrm{f}_1(\w)}$ one could take
into account, if to recall the
 transformation freedom connected with the basis renormalization (\ref{auf}). Then
the matrix elements will be changed in this way $c_1(\u,\w)\to
[s/\bar{s}]c_1(\u,\w)$, $\bar{c}_1(\u,w)\to
[\bar{s}/s]c_1(\u,\w)$, $c_3(\u,\w)\to [t/\bar{t}]c_3(\u,\w)$,
$c_2(\u,\w)\to [s t/(\bar{s}\bar{t})]c_2(\u,\w)$ and
$\bar{c}_2(\u,\w)\to [\bar{s}\bar{t}/(s t)]\bar{c}_2(\u,\w)$,
forming the "eight-parametric" matrix
$R^{\alpha\bar{\alpha}}_{33}(u,w;p,q;\bar{p},\bar{q};
s/\bar{s};t/\bar{t})$.

In the scheme of the Quantum Inverse Scattering Problem \cite{FT}
the structure of the one-dimensional quantum spin Hamiltonian
corresponding to the $R_{33}$-matrix
 can be seen presenting the $\check{R}_{33}$ in the %following
operator form, preliminary doing the notations
\bea e^{\pm}=\frac{S^z(S^z\pm 1)}{2}, \quad e^0=I-(S^z)^2,\quad
S^{ik}=S^i S^k,\; i,k=+,-,z, \ena
 where $I$ is the unite operator and
 $S^{\pm}=\sqrt{2}J^{\pm},\;S^z=J^z$.
$J^i$ are the normalized $3\times 3$ spin-1
generators of the $sl_2$-algebra, $J^+=\frac{1}{\sqrt{2}}{\scriptsize\left(\ba{ccc}&1&\\
&&1\\&&\ea\right)}$,
$J^-=\frac{1}{\sqrt{2}}{\scriptsize\left(\ba{ccc}&&\\1
&&\\&1&\ea\right)}$ and $J^z={\scriptsize\left(\ba{ccc}1&&\\
&&\\&&-1\ea\right)}$. Then the $R$-matrix has the general
structure (below we are omitting the arguments $\u,\w$ of the
functions)
\bea\label{hr-33} &\check{R}_{33}=a_1 e^+\otimes e^++a_2
e^0\otimes e^0 +a_3 e^-\otimes e^-+&\\&b_1 S^{z+}\!\otimes
S^{-z}\!+ \bar{b}_1 S^{-z}\!\otimes S^{z+}\!+b_2 S^{+z}\!\otimes
S^{z-}\!+ \bar{b}_2 S^{z-}\!\otimes S^{+z}+b_3 S^{++}\!\otimes
S^{--} \!+\bar{b}_3 S^{--}\!\otimes S^{++}\!+ \nn &\\&c_1
e^0\otimes e^++\bar{c}_1 e^+\otimes e^0+c_2 e^-\otimes
e^0+\bar{c}_2 e^0\otimes e^-+ c_3 e^-\otimes e^++\bar{c}_3
e^+\otimes e^-.&\nn \ena
Let us adopt a convention that we have one kind spectral parameter
$\u=u$, $f(\u)=u+1$ and two arbitrary functions
$\mathbf{c}_{1,3}(u)$.
 Then expanding by the variable $w$ the $\check{R}_{33}(u,w)$-matrix near the point $u$,
the linear terms in the expansion will correspond to the local
cell terms $H_{i,i+1}$ of the respective spin-Hamiltonian  with
the interactions between the nearest-neighbors spins (acting on
the spaces $V_{i}\otimes V_{i+1}$): $\check{R}_{33}\approx
I\otimes I+(w-u) H_{i,i+1}$. For the simplicity we can take
$\mathbf{c}_{i}$ to be constant -
$\mathbf{c}_1(u)=e^{\varepsilon_1/2}$ and
$\mathbf{c}_3(u)=e^{\varepsilon_3/2}$, $\varepsilon_i$ are
numbers. The extension for the general family of the Hamiltonian
operators with arbitrary functions $\mathbf{c}_{1,3}(u)$ would be
obvious.
\bea &\check{R}_{33}(u,w)=I\otimes
I+\frac{w-u}{1+u(1-x_f)}P_{\alpha,\bar{\alpha},\gamma,\varepsilon_i}+
\frac{w-u}{1+u(1-x_f)}(x_f-1)\bar{P}_{\alpha,\bar{\alpha},\gamma,\varepsilon_i}+O(w-u),&\\&
P_{\alpha,\bar{\alpha},\gamma,\varepsilon_i}= e^+\otimes
e^++\alpha \; e^0\otimes e^0 +\bar{\alpha}\; e^-\otimes e^-+
e^{\varepsilon_1} S^{z+}\!\otimes S^{-z}+ e^{-\varepsilon_1}
S^{-z}\!\otimes S^{z+}+\nn &\\&
{(e^{\varepsilon_3-\varepsilon_1})}{\gamma}  S^{+z}\!\otimes
S^{z-}+ \frac{e^{\varepsilon_1-\varepsilon_3}}{\gamma}
S^{z-}\!\otimes S^{+z}+ e^{\varepsilon_3} S^{++}\!\otimes S^{--}
+e^{-\varepsilon_3} S^{--}\!\otimes S^{++}, &\\&
\bar{P}_{\alpha,\bar{\alpha},\gamma,\varepsilon_i}= e^+\otimes
e^++\frac{1+\alpha}{2} \; e^0\otimes e^0 +
\frac{1+\bar{\alpha}}{2}\; e^-\otimes e^- +\frac{ e^0\otimes e^++
e^+\otimes e^0+
 e^-\otimes e^++ e^+\otimes e^-}{2}+ e^0\otimes e^-+&\nn\\&
2\left(e^{\varepsilon_3} S^{++}\!\otimes S^{--}+e^{-\varepsilon_1}
S^{-z}\!\otimes S^{z+}+
{(e^{\varepsilon_3-\varepsilon_1})}{\gamma}  S^{+z}\!\otimes
S^{z-}\right). \label{pxf} & \ena
The operators $P_{\alpha,\bar{\alpha},\gamma,\varepsilon_i}$ at
the values $\gamma=\pm 1,\;\varepsilon_i=\pi n_i$ ($n_i$ are
integers)
 just correspond to the permutation operators - the signs
$\alpha,\bar{\alpha}=\pm 1$ and $\gamma,\; e^{\varepsilon_i}$ are
responsible to the different gradings of the spaces. Such that the
ordinary non graded case $\alpha=\bar{\alpha}=1$,
$\gamma=1,\;\varepsilon_i=0$ corresponds to the  permutation
operator of the $sl(2)$-invariant spin-$1$ spaces. The values
$\alpha=-1,\; \bar{\alpha}=1$, $\gamma=1,\;\varepsilon_i=0$
correspond to the $osp(1|2)$-invariant case with the fundamental
three dimensional spin-$1/2$ representation spaces with two even
and one odd parity vectors,. When $\alpha=-1,\; \bar{\alpha}=1$,
$\gamma=1,\;e^{\varepsilon_1}=-1, \;e^{\varepsilon_3}=1 $, the
corresponding  operator, multiplied by a minus sign, is the
permutation acting on the spaces with two odd and one even parity
vectors.
  For the three
dimensional solutions of the mentioned algebras and their quantum
extensions see the papers \cite{TsBj,Saleur,PPKNYR,ShK,Isaev}, and
it is interesting to mention that taking into account the
correspondence between the representations of the quantum algebras
$osp_q(1|2)$ and $sl_{i\sqrt{q}}(2)$, the mentioned graded
matrices can be obtained from the  $sl_t(2)$-invariant matrices at
$t=1,i$.
  The next summand with $\bar{P}$ (\ref{pxf}) exists when $x_f\neq 1$ and spoils the
 mentioned symmetries even  at the points $\gamma=1,\;
 \varepsilon_i=0$. We must menton once again, that in the course
 of the solution we have taken $c_2\neq \bar{c}_2$, and this
 choice induces the mentioned Hamiltonian. The solutions with $c_1\neq
 \bar{c}_1$ or  $c_3\neq \bar{c}_3$ would bring to slight changes in the
 structure of the asymmetric part $\bar{P}$, conditioned by the
 appropriate interchanges of the coefficients in (\ref{hr-33}).

\section{Summary}

%  We can summarize the results obtained in the paper in the following points.
  In this paper
  we have completed the list of the colored $R_{22}$ solutions by multi-parametric
  free-fermionic solutions.
 Colored $R_{33}$-matrices are obtained for the matrices with $15$ non-zero entries
  The given approach for the solving the multi-parametric YBE,
   being very simple and clear-cut, gives an opportunity
  to find colored YBE solutions for higher dimensional matrices.
  It is shown, that the number of the possible extra colors, on which
  the $R$-matrix can be dependent,
    is restricted: %at the beginning
initially the number of
  the elementary functions (the matrix elements of
$R(\u,\mathbf{0})$ taken at the normalization point $\mathbf{0}$)
 is defined from the number of the non-zero elements, and then the set
 of the independent relations following from YBE
 imposed on the
  elementary functions determines the number of the
arbitrary  functions (or the number of the possible
  colors) existing in the solutions.
  The solutions for YBE $(^\ast)$ $K(\u,\w,\mathbf{0})=0\;(
  K(\u,\mathbf{0},\w)=0,\;K(\mathbf{0},\u,\w)=0)$ are sufficient
  to solve the whole set of the
  YBE, as the taken
  point $\mathbf{0}$ is chosen arbitrarily.

 The discussed multi-parametric $R$-matrices being the solutions of the
   Yang-Baxter equations, can have usage and treatment
   in all areas of the theoretical and mathematical physics,
   where YBE are involved - integrable models, high energy
   physics, quantum groups,
    quantum information theory, statistical physics. We have
    seen that the four-parametric $4\times 4$-solutions presented in the
    section 2.3.3 have their interpretation as  intertwiner
    matrices
    in the theory of the quantum algebra \cite{shkh} (i.e. all free-fermionic
    solutions can be presented
    as $sl_q(2)$-invariant
    matrices at $q=i$), and the colored
    parameters are the characteristics of the representations of
    the
    quantum algebra. For the cases with higher dimensions  also
    one could expect the existence of respective underlying symmetry.
    In the framework of the Algebraic Bethe Ansatz the multi-parametric solutions
  bring to the richer families of integrable Hamiltonians.  The
  cases when there are more than one pair of the spectral parameters
  with "difference" property (which
  can be interpreted as "rapidities"), as it was
  in (\ref{raa}, \ref{ybezuv}),  are of particular interest.

The all obtained solutions are unitary in the following sense
 \bea
 \check{R}(\u,\w)\check{R}(\w,\u)\approx
 \mathbf{I}.
 \ena
 This relation can be proved for all the cases,
 using only the symmetry property of the functions $f_i(\u,\w)$
 under the interchange of the variables $\u$ and $\w$, and the
 compatibility conditions of the Yang-Baxter equations.
 Particularly, the unitarity relation for the $R_{22}$-matrix
\bea
  a_1(\u,\w)a_1(\w,\u)+d(\u,\w)d(\w,\u)=1+b_1(\u,\w)b_2(\w,\u),
 \ena
for the free-fermionic solutions just means the free-fermionic
property. If to take the usual unitarity condition for the
matrices,
 \bea
 \check{R}^{+}(\u,\w)=\check{R}^{-1}(\u,\w), \label{unitar}\ena
  then we shall have
 some restrictions on the solutions, but the number of the arbitrary
 functions does not changed, as the relation (\ref{unitar}) means
\bea a_1(\u)^{+}=a_2(\u),\quad b_1^{+}(\u)=-b_2(\u),\quad
d^{+}(\u)=-d(\u). \ena
Here there are five relations on the ten functions
$\mathrm{Re}[f_i(\u)],\; \mathrm{Im}[f_i(\u)]$ and again we have
five elementary functions: $\mathrm{Re}[a_1(\u)](=\mathrm{Re}[a_2(\u)])$,
$\mathrm{Re}[b_1(\u)](=-\mathrm{Re}[b_2(\u)])$,
$\mathrm{Im}[a_1(\u)](=-\mathrm{Im}[a_2(\u)])$,
$\mathrm{Im}[b_1(\u)](=\mathrm{Im}[b_2(\u)])$ and
$\mathrm{Im}[d(\u)]\;(\mathrm{Re}[d(\u)]=0)$, with further (in
general three) relations on them.

\paragraph{Note.}  We could investigate in the same manner the cases with the conditions
$f_1=0,\; f_2\neq 0$ or $f_2=0,\; f_1\neq 0$ with $f=b,d$ (the
situations with $f=a,c$ would spoil the condition (\ref{unite})
and bring to rather trivial solutions). As we have learnt from
\cite{SSh,shkh} such cases can include separate solutions,
together with the limit cases of the general solutions which one
can obtain  after taking the appropriate limits ($f_i\to 0$, $i=1$
or $i=2$), and however all the solutions have functional
dependence (i.e. existence of the definite number of arbitrary
functions, or in the particular cases, elliptic, trigonometric
and rational future of the solutions) similar to the case of
general solutions.

\paragraph{Acknowledgements.} %The work is partly supported by an
%Armenian grant 11-1c028 and by ANSEF grant matph-2908.

This work was supported by State Committee Science MES RA, in
frame of the research project. The work was made possible in part
by a research grant from the Armenian National Science and
Educational Fund (ANSEF) based in New York (USA).

%\addtocounter{section}{0}\setcounter{equation}{0}

\setcounter{section}{0}
\renewcommand{\thesection}{\Alph{section}}
\Alph{section}
\section{Appendix}
\renewcommand{\theequation}{\Alph{section}.\arabic{equation}}
\addtocounter{section}{0}\setcounter{equation}{0}

\subsection*{The independent equations in the set of YBE with the conditions
 $d(\u)\neq 0$ and the derivation of a %:detailed discussion of the
 solution  %process
 at $a_1'(\mathbf{0})=-a_2'(\mathbf{0})$
}
 %\addtocounter{section}{0}\setcounter{equation}{0}

%\renewcommand{\theequation}{\Alph{section}.\arabic{equation}}
%\addtocounter{section}{0}\setcounter{equation}{0}

 Here we discuss the set of the YBE for the case with the
 conditions $b_1(\u)a_1(\u)=b_2(\u)a_2(\u)$,
 $a_1'(\mathbf{0})=-a_2'(\mathbf{0})$. As in the general case, here there are six independent equations in the set of YBE.
 After performing the following notations

\bea f_1(\u,\v)=f_{12},\quad f_2(\u,\v)=f_{x12},\quad
 f_1(\u,\w)=f_{13},\quad f_2(\u,\v)=f_{x12},\nn\\
 f_1(\v,\w)=f_{23},\quad f_2(\v,\w)=f_{x23},\quad f=a,b,d.\ena
 the independent equations look like as
\bea \label{ybe1c}&a_{12}a_{23}-a_{13}-b_{23}b_{x12}+a_{x13}d_{12}d_{23}=0,&\\
&a_{12}d_{13}-a_{x12}d_{23}-a_{13}a_{23}d_{12}+b_{x13}b_{x23}d_{12}=0,&\\
&a_{12}b_{13}-a_{13}b_{12}-b_{23}+b_{x23}d_{12}d_{13}=0,&\\
&a_{23}b_{x13}-b_{x12}-a_{13}b_{x23}+b_{12}d_{13}d_{23}=0,&\\
&a_{23}d_{13}-a_{x23}d_{12}-a_{12}a_{13}d_{23}+b_{12}b_{13}d_{23}=0,&\\
&b_{x13}d_{12}-b_{13}d_{23}+a_{12}b_{23}d_{13}-a_{23}b_{x12}d_{13}=0.\label{ybe1e}
\ena

  The remaining equations can be obtained from these
 ones doing some changes of the arguments, taking into account that
\bea a_1(\u,\w)=a_2(\w,\u),\quad b_i(\u,\w)=-b_i(\u,\w),\quad
d(\u,\w)=-d(\w,\u) \label{condit} \ena

Let us prove that under the mentioned conditions the relations
(\ref{a1b1}, \ref{a2b2}, \ref{abxf}) imposed on the functions
$f_i(\u)$, $f=a,b,d$ are enough to solve the whole set of YBE. The
equations appear to be large and rather complicated and for
simplifying the evaluation process we could suggest the following
%way to proceed after doing some
 reparameterizations. Let us
introduce new functions $f(\u)$ and $g(\u)$, such that
\bea a_1(\u)=g(\u)a_2(\u) \qquad b_1(\u)=f(\u)a_2(\u).\ena
 We can
parameterize them by means of the function $d(\u)$ and two
constants $f_0,\; x_f$ %$f_0=x_z$ and $x_f=x_0/x_z$.
\bea\label{fg}
&f(\u)=f_0\left(f_g(\u)-\sqrt{f_g(\u)^2-\frac{1}{f_0^2}}\right),\quad
g(\u)=\frac{2 d(\u)}{1+d(\u)^2}\left(x_f+f_g(\u)\right)&\\
&f_g(\u)=\sqrt{x_f^2+\frac{(1+d(\u)^2)^2}{4 d(\u)^2}},\quad
a_2(\u)=\sqrt{\frac{1+d(\u)^2}{g(u)(1+f(\u)^2)}}.& \ena
The two parametric functions followed from (\ref{abd}) can be
written in rather compact formulas
\bea &a_1(\u,\w)=a_2(\u)a_2(\w)\left[1+f(\u)f(\w)\right]\mathcal{A}_1(\u,\w)&\\
&a_2(\u,\w)=a_2(\u)a_2(\w)\left[1+f(\u)f(\w)\right]{A}_2(\u,\w)&\\
&b_1(\u,\w)=a_2(\u)a_2(\w)\left[f(\u)-f(\w)\right]\mathcal{B}_1(\u,\w)&\\
&b_2(\u,\w)=a_2(\u)a_2(\w)\left[f(\u)-f(\w)\right]\mathcal{B}_1(\u,\w)&\\
&d(\u,\w)=\frac{f(\u)-f(\w)}{f(\u)+f(\w)}\mathcal{D}(\u,\w).& \ena
Here we have introduced the following functions, which can be
parameterized only by the functions $g(\u),\;d(\u)$ (from the
equations (\ref{fg}) it follows
$f_g(\u)=g(\u)\frac{(1+d(\u)^2)}{2d(\u)}-x_f$)
\bea
\mathcal{A}_1(\u,\w)=\frac{d(\w)+d(\u)g(\u)g(\w)}{2d(\u)d(\w)[f_g(\u)+f_g(\w)]},\quad
A_2(\u,\w)=A_1(\w,\u)\\
\mathcal{B}_1(\u,\w)=\frac{1+d(\u)d(\w)g(\u)g(\w)}{1-d(\u)^2
d(\w)^2},\quad
\mathcal{B}_2(\u,\w)=\frac{d(\u)d(\w)+g(\u)g(\w)}{1-d(\u)^2
d(\w)^2}\\
\mathcal{D}(\u,\w)=\frac{\left(d(\u)-d(\w)\right)\left(1-g(\u)g(\w)\right)}{
\left(1+d(\u)d(\w)\right)\left(g(\u)-g(\w)\right)}
 \ena
 We see that the functions factorize into two parts which
contain the function $f(\u)$ and the functions $d(\u),\; g(\u)$.
In the equations we can expand the relations on the series in
terms of the constant $f_0$ and the function-factors
$f_{x1}=\sqrt{f_g(\u)^2-\frac{1}{f_0^2}}$,
$f_{x2}=\sqrt{f_g(\v)^2-\frac{1}{f_0^2}}$,
$f_{x3}=\sqrt{f_g(\w)^2-\frac{1}{f_0^2}}$, as they meet only in
the functions $f(\u),\;f(\v),\;f(\w)$. %and the remaining functions
%can be parameterized by familiar elliptic parameterizations.
 Also
one can notify that the functions $a_2(\u)$ either have been
factorized from the equations or met there in the quadratic form
$a_2(\u)^2=\frac{a_f(\u)}{1+f(\u)^2}$,
 $a_f(\u)=\frac{1+d(\u)^2}{g(\u)}$, which allows us to escape the
double square roots in the equations.

As example, let us represent the investigation of the first
equation (\ref{ybe1c}). After the mentioned expansion we find,
that there are only two following equations, which one has to
prove and which can be done after some  not so complicated
calculations.
\bea
A_1(\u,\w)-a_f(\v)B_1(\v,\w)B_2(\u,\w)+D(\u,\v)D(\v,\w)A_2(\u,\w)=0,\qquad\\
2A_1(\u,\w)f(\v)[f_g(\u)+f_g(\w)]-
a_f(\u)\Big(B_1(\v,\w)B_2(\u,\v)[f_g(\u)-f_g(\v)]
[f_g(\v)-f_g(\w)]+\nn\\
A_1(\u,\v)A_1(v,\u)[f_g(\u)+f_g(\v)][f_g(\v)+f_g(\w)]\Big)=0.\qquad
\ena
%
%parenthesis

{\small
\subsection*{ The main $4\times 4$ solutions to YBE
 obtained  heretofore%, observed  and
%classified in \cite{Baxt,Felder,BVS}, two-parametric solution in
%\cite{SSh,shkh}, solutions with arbitrary function \cite{shkh}
}
 %\addtocounter{section}{0}\setcounter{equation}{0}
Here we are presenting in the apparent matrix form the already
known solutions,
 observed  and
classified in \cite{Baxt,Felder,BVS} and  two-parametric solutions
in \cite{SSh,shkh}.
%\subsubsection*{The solutions with the difference property}
\paragraph{The eight-vertex solution} with elliptic parameterization is
\cite{Baxt}

\bea\label{ybezi}
 R_{xyz}(u)\!=\!\!\left(\!\!\ba{cccc}
  \frac{\mbox{sn}[{u+\lambda},\;k]}{\mbox{sn}[\lambda,\;k]}&0&0&{
  e^{\gamma/2}k\;\mbox{sn}[{\lambda+u},k]\mbox{sn}[u,k]}\\
0&\frac{\mbox{sn}[u,\;k]}{\mbox{sn}[\lambda,\;k]}&1&0\\
0&1&\frac{\mbox{sn}[u,\;k]}{\mbox{sn}[\lambda,\;k]}&0\\
{e^{-\gamma/2}k\;\mbox{sn}[{\lambda+u},k]\mbox{sn}[u,k]}
&0&0&\frac{\mbox{sn}[{u+\lambda},\;k]}{\mbox{sn}[\lambda,\;k]}
\ea\!\!\right).
 \ena

The following relation takes place
$\frac{a^2(u)+b^2(u)-c^2(u)-d^2(u)}{a(u)b(u)}=
2\mbox{cn}[\lambda,k]\mbox{dn}[\lambda,k]$.

Free fermionic case corresponds to the relation
$\mbox{cn}[\lambda,k]\mbox{dn}[\lambda,k]=0$. The $XY$-model
corresponds to $\lambda=K(k)$.

All other solutions presented below have free-fermionic property.

\paragraph{The non-homogeneous solution $a_1(\u)=a_2(\u)$,
$b_1(\u)=-b_2(\u)$} corresponds to \cite{SSh}

\bea\label{ybezuv}
 \tilde{R}(u;v)\!=\!\!\left(\!\!\ba{cccc}
 \cosh{[u]}&0&0&{
  e^{\gamma/2}\;\sin{[v]}}\\
0&\sinh{[u]}&\cos{[v]}&0\\
0&\cos{[v]}&-\sinh{[u]}&0\\
{e^{-\gamma/2}\sin{[v]}} &0&0&\cosh{[u]}\ea\!\!\right).
 \ena
 and satisfies to the YB equations (\ref{ybeuu'}).
 %
%\bea \sum_{j_1,j_2,j_3} R_{i_1 i_2}^{j_1 j_2}(u;v)R_{j_1 i_3}^{k_1
%j_3}(u+w;v+y)R_{j_2 j_3}^{k_2
%k_3}(w;y)=\qquad\qquad\qquad\nn\\\qquad\qquad\qquad\sum_{j_1,j_2,j_3}
%R_{i_2 i_3}^{j_2 j_3}(w;y)R_{i_1 j_3}^{j_1 k_3}(u+w;v+y)R_{j_1
%j_2}^{k_1 k_2}(u;v).\label{ybeuv}\ena
%

\paragraph{The non-homogeneous one-parametric solution $a_1(\u)\neq a_2(\u)$,
$b_1(\u)=b_2(\u)$} corresponds to \cite{SSh}
\bea\label{rxydc}
 R_I(u)\!=\!\!\left(\!\!\!\ba{cccc}
\frac{\mbox{dn[u,k]}}{\mbox{cn[u,k]}}\pm\frac{\mbox{dn}[\mbox{u}_0,\mbox{k}]\mbox{sn[u,k]}}{\mbox{sn}[\mbox{u}_0,\mbox{k}]}
 &0&0& e^{\gamma/2}\frac{\mbox{dn[u,k]}\mbox{sn[u,k]}}{\mbox{cn[u,k]}} \\
0&\frac{\mbox{sn[u,k]}}{\mbox{sn}[\mbox{u}_0,\mbox{k}]}&1&0\\
0&1
&\frac{\mbox{sn[u,k]}}{\mbox{sn}[\mbox{u}_0,\mbox{k}]}&0\\
e^{-\gamma/2}\frac{\mbox{dn[u,k]}\mbox{sn[u,k]}}{\mbox{cn[u,k]}}&0&0&
\frac{\mbox{dn[u,k]}}{\mbox{cn[u,k]}}\mp\frac{\mbox{dn}[\mbox{u}_0,\mbox{k}]\mbox{sn[u,k]}}{\mbox{sn}[\mbox{u}_0,\mbox{k}]}
\ea\!\!\!\right)\!.
 \ena
%
%
%\bea\label{rxycd}
% R_{2}(u,k)\!=\!\!
% \left(\!\!\!\ba{cccc}
%\frac{\mbox{cn[u,k]}}{\mbox{dn[u,k]}}\pm\frac{\mbox{cn}[\mbox{u}_0,\mbox{k}]\mbox{sn[u,k]}}{\mbox{sn}[\mbox{u}_0,\mbox{k}]}
% &0&0& e^{\gamma/2}k\frac{\mbox{cn[u,k]}\mbox{sn[u,k]}}{\mbox{dn[u,k]}} \\
%0&\frac{\mbox{sn[u,k]}}{\mbox{sn}[\mbox{u}_0,\mbox{k}]}&1&0\\
%0&1
%&\frac{\mbox{sn[u,k]}}{\mbox{sn}[\mbox{u}_0,\mbox{k}]}&0\\
%e^{-\gamma/2}k\frac{
%\mbox{cn[u,k]}\mbox{sn[u,k]}}{\mbox{dn[u,k]}}&0&0&
%\frac{\mbox{cn[u,k]}}{\mbox{dn[u,k]}}\mp\frac{\mbox{cn}
%[\mbox{u}_0,\mbox{k}]\mbox{sn[u,k]}}{\mbox{sn}[\mbox{u}_0,\mbox{k}]}
%\ea\!\!\!\right)\!.
% \ena
%

If $u_0=K(k)$, where $K$ is the complete elliptic integral of the
first kind with the module $k$, then the matrix $R_I(u)$
corresponds exactly to the $R$-matrix, which we have found in
\cite{ShAS} for $2d$ Ising Model. After %The solution, with equivalent elliptic parameterization,
 %connected  with the presented one by the
so called "transformations of the first degree" of the elliptic
functions, this solution  at $u_0=K(k)$ corresponds to the
$XY$-model matrix.

\paragraph{The three-parametric (colored) solution in \cite{BVS,Felder}.}
Defining the function $\mbox{e}[\mathrm{u},\mathrm{k}]$ as
$\mbox{e}[\mathrm{u},\mathrm{k}]=\mbox{cn}[\mathrm{u},\mathrm{k}]+i
\mbox{sn}[\mathrm{u},\mathrm{k}]$ the matrix is written as
\bea \label{color} R(\mathrm{u};p,q)=\qquad\\\nn
{\scriptsize{\left(\!\!\!\ba{cccc}1-\mbox{e}[\mathrm{u},\mathrm{k}]pq&0&0&\frac{\mathrm{k}\sqrt{(1-p^2)(1-q^2)}(1+\mbox{e}[\mathrm{u},\mathrm{k}])
\mbox{sn}[\frac{\mathrm{u}}{2},\mathrm{k}]}{2}\\0&q\!-\!p\mbox{e}[\mathrm{u},\mathrm{k}]
&\frac{i\sqrt{(1-p^2)(1-q^2)}(1-\mbox{e}[\mathrm{u},\mathrm{k}])}{2\mbox{sn}[\frac{\mathrm{u}}{2},\mathrm{k}]}&0\\0&\frac{i\sqrt{(1-p^2)(1-q^2)}(1-\mbox{e}[\mathrm{u},\mathrm{k}])}{2\mbox{sn}[\frac{\mathrm{u}}{2},\mathrm{k}]}
&p\!-\!q\mbox{e}[\mathrm{u},\mathrm{k}]
&0\\
\frac{\mathrm{k}\sqrt{(1-p^2)(1-q^2)}(1+\mbox{e}[\mathrm{u},\mathrm{k}])\mbox{sn}[\frac{\mathrm{u}}{2},\mathrm{k}]}{2}
&0&0&\mbox{e}[\mathrm{u},\mathrm{k}]- p q \ea\!\!\!\right)}}.\ena
The second solution given in the subsection 2.3.2 corresponds to
this case after transformation $\mathrm{k}\to 1/k=i/x_f$. Taking
the symmetric parameterization $p=\mbox{e}[\psi_1,\mathrm{k}]$,
$q= \mbox{e}[\psi_2,\mathrm{k}]$ one can write the matrix as
function $R(\mathrm{u};\psi_1,\psi_2)$. }

\paragraph{$sl_q(2)$-invariant solution defined on the
two-dimensional cyclic irreps at $q=i$.}

The $sl_q(2)$-invariant $4\times 4$ $R$-matrices at $q=i$ have the
free-fermionic property: the solutions defined on the spin-irreps
correspond to the trigonometric $XX$ limit of the
$R_{xyz}$-matrix (\ref{ybezi}) when $d_i=0$, and the first
observed solution \cite{RAM} on the cyclic (semi-cyclic,
nilpotent) irreps corresponds to the same $d_i=0$ limit of the
inhomogeneous matrix (\ref{rxydc}). The solutions defined on the
cyclic irreps with special degenerated cases are described in
\cite{shkh} (sections 4.3, 4.4), where $c_i \cosh {\varepsilon_j}=
c_j\cosh{\varepsilon_i}=0$, with $c_{i,j},\; exp\{\varepsilon_{i,j}\}$
 being the eigenvalues of the Casimir
 operators $c$ and $k^2$ from the extended center of the quantum algebra. Besides of the
special trigonometric limits,  now with $d_i\neq 0$, of the
matrices (\ref{ybezi}, \ref{rxydc}, \ref{ybezuv}, \ref{color}),
the solutions presented there include also general solutions with
two constants  which do not admit difference property. The
mentioned solutions can coincide with the  solutions obtained here
in the subsection 2.3.3 having
  two arbitrary functions
and two arbitrary constants, after fixing the arbitrary functions
(say $d(\u),\; f_g(\u)$) to have  the following dependence on the
exponential function $e^{\varepsilon}$ and the arbitrary function
$\bar{h}(\varepsilon)$ established in  \cite{shkh}
\bea
d(\u)=\bar{h}(\varepsilon),\\
f_g(\u)=i\frac{[\bar{f}(\varepsilon)e^\varepsilon-1]+
\bar{h}(\varepsilon)[\bar{f}(\varepsilon)-g_0
e^\varepsilon]}{[e^\varepsilon+\bar{f}(\varepsilon)]-
\bar{h}(\varepsilon)[e^\varepsilon \bar{f}(\varepsilon)+g_0]},\\
2 h_0 \bar{f}(\varepsilon)\bar{h}(\varepsilon)=
1-\bar{f}(\varepsilon)^2+\bar{h}(\varepsilon)^2[g_0^2-\bar{f}(\varepsilon)^2].
\ena
The function $\bar{f}(\varepsilon)$ is related to $\varepsilon$
and $\bar{h}(\varepsilon)$ by means of the last relation, $h_0$
and $g_0$ are  arbitrary numbers \cite{shkh},
 which can be expressed by the constant parameters $f_0$ and $x_f$.

%\begin{figure}[h]
%\unitlength=5pt
%\unitlength=20pt
%\hspace{-3cm}\vspace{-7cm}
%\includegraphics[width=150mm,angle=0,clip]{aghyusak.eps}
%\caption{A classification of the solutions}%\begin{figure}[h]
%\label{fig1}%\vspace{0.5cm}
%\end{figure}


\begin{thebibliography}{99}

\bibitem{Y} C.~N.~Yang, Phys. Rev. Lett. {\bf 19} (1967) 1312.

\bibitem{Baxt78} R.~J.~ Baxter, {\it Solvable eight-vertex model on an arbitrary
planar lattice}, Proc. Roy. Soc. {\bf 289 A} (1978) 2526-47.

\bibitem{Ons} A.~Onsager, %{\it Crystal statistics I: a two-dimensional
%model with an order-disorder transition} -
Phys.Rev. {\bf 65} (1944) 117-49.

\bibitem{FT} L.~D.~ Faddeev,  L.~A.~Takhtajan, {\it The quantum inverse problem method
and the XYZ Heisenberg model}, Usp. Mat. Nauk {\bf 34} (1979) 13-
194.

\bibitem{Baxt} R.~J.~ Baxter, {\it Exactly solvable models in Statistical Mechanics
}, Academic Press, London (1982).

\bibitem{FTS} L.~D.~ Faddeev, E~K.~ Sklyanin and L.~A.~Takhtajan,
Theor. Math. Phys. {\bf 40} (1979) 194.

\bibitem{Z} A~B.~ Zamolodchikov and Al.~B.~ Zamolodchikov, Ann.
Phys. {120} (1979) 253.

\bibitem{Z1} A.~B.~ Zamolodchikov, Sov. Sci. Rev. {\bf A2} (1980)
1.

\bibitem{grs} C.~Gomez, M.~ Ruiz-Altaba, G.~Sierra, {\it Quantum groups
in two-dimensional physics}, Cambridge, University Press (1993).


\bibitem{ShAS} Sh.~Khachatryan, A.~Sedrakyan, %{\textit{Charactersistics
%of the 2D Ising model from fermionic realization}} -
ArXiv: 0712.0273v1, Phys. Rev. {\bf B 80} (2009)  125128.

\bibitem{WuChan} Ch.%ungpeng
~ Fan, F.~ Y.~ Wu, %{\textit{General Lattice Model
%of phase Transitions}}
Phys. Rev. {\bf B 2} (1970) 723-733.



\bibitem{Korn} G.~H.~Korn, T.~M.~Korn, {\it Mathematical Handbook}
- the russian translation of the second completed edition, Moscow,
Nauka (1973).

\bibitem{Felder} B.~U.~Felderhof, Physica {\bf 66} (1973) 279.

\bibitem{BVS}  V.~V.~Bazhanov, Yu.~G.~Stroganov, %{\it Hidden symmetry of
%free fermion model}
- Teor. Mat. Fiz. {\bf 62} (1985) 377.

\bibitem{AuP} Jacques H.H. Perk $\&$
 Helen Au-Yang %{\it YANG–BAXTER EQUATIONS}
  {\sl{Phys. Lett.}} {\bf{A 123}} 219
(1987).

\bibitem{AuPY}  R.~J.~Baxter,
J.~H.~H.~Perk, H.~Au-Yang, {\sl{Phys. Lett.}} {\bf{A 128}} 138-147
(1988).

 \bibitem{SSh} Sh.~Khachatryan, A.~Sedrakyan, ArXiv:1208.4339, J. Stat. Phys.
 150 (2013) 130-155.

\bibitem{shkh} D.~Karakhanyan, Sh.~Khachatryan,
 %{\textit{YBE solutions with $sl_q(2)$ symmetry at roots of unity: cylic irreps.}} -
 ArXiv: 1203.6528v1, %doi: 10.1016/ j. nuclphysb. 2012.11.003.
 Nucl. Phys. {\bf B} 868 [PM] (2013) 328-349.

%\bibitem{sha} Sh.~Khachatryan, article in preparation.

\bibitem{RKS} T.~Deguchi, Y.~Akutsu, J. Phys. Soc. Jpn.
{\bf 62} No.1 (1993) 19-35; T.~Deguchi, Y.~Akutsu, J. Phys. Soc.
Jpn. {\bf 60} No.12 (1991) 4051-4059.

\bibitem{RAM} M.~ Ruiz-Altaba, {\sl{Phys. Lett.}}
{\bf{B 279}} 326-332 (1992);


\bibitem{Wang} Shi-Kun Wang, ArXiv:q-alg/9712020.

\bibitem{GU} G.~Uimin, JETP {\bf 12} (1970).

 \bibitem{JKL} J.~K.~Lai, Journal of Math. Physics {\bf 15} (1974)
1675.

\bibitem{BS} B.~Sutherland, Phys. Rev. {\bf B 12} (1975) 3795.

\bibitem{PerkSch}
 J.~H.~H.~Perk and C.~L.~Schultz,
%  "New families of commuting transfer matrices in $q$-state vertex models,"
  Phys. Lett. {\bf A 84} (1981) 407-410;
  J.~H.~H.~Perk and C.~L.~Schultz,
%  "Families of commuting transfer matrices in $q$-state vertex models"
  in "Non-linear integrable systems - classical theory and quantum
  theory",
  Proceedings of RIMS Symposium organized by M. Sato, Kyoto, Japan,
  13-16 May 1981, eds. M. Jimbo and T. Miwa,
   World Scientific, Singapore, (1983), pp. 135-152;
  C.~L.~Schultz, "Commuting Transfer Matrices of the Closed Colored
  String Model in Two Dimensions," Ph.D Thesis, State University of
  New York at Stony Brook, December 1981, Chapter II, pp. 33-61.

\bibitem{PerkAu-Yang}   J.~H.~H.~Perk and H.~Au-Yang,
  "Yang-Baxter Equation," in "Encyclopedia of Mathematical Physics,"
  eds. J.-P. Fran\c{c}oise, G.L. Naber and Tsou S.T.,
  Vol. 5, (Oxford: Elsevier Science, May 2006, ISBN 978-0-1251-2666-3),
  pp. 465-473, %[See eqs. [20] and [21] and surrounding text.]
   arXiv:math-ph/0606053.

\bibitem{33}  R.~A.~Pimenta,
M.~J.~Martins, Journal of Physics {\bf A}: Math. and Theor. {\bf
44}, Issue 8 (2011) 085205.

\bibitem{33osp} R.~A.~Pimenta,
M.~J.~Martins
 and M. Zuparic,
%{\it The factorized F-matrices for arbitrary U(1)(N-1) integrable vertex models}
 Nucl. Phys.
  {\bf B} {\bf
859}, Issue 2 (2012) 207–260.


\bibitem{TsBj} H.~M.~Babujian, Physics Letters {\bf A 90} (1982)
479; Hratchya~M.~Babujian, A.~M.~Tsvelik, Nucl.Phys. {\bf B 265}
(1986) 24.

\bibitem{Saleur} H.~Saleur, Nucl. Phys. {\bf B 336} (1990) 363.

\bibitem{PPKNYR}
P.~P.~Kulish, N.Y. Reshetikhin, Lett. Math. Phys. {\bf 18} (1989)
 143.

\bibitem{ShK}
D.~Karakhanyan, Sh.~Khachatryan, Nuclear Physics {\bf B 808} [FS]
(2009) 525-545.

\bibitem{Isaev} A.~O.~Isaev, J. Phys. {\bf A 28}, Math. Gen.
(1996) 6903.


\end{thebibliography}
\end{document}